\documentclass{ar-astro2e}

\usepackage[margin=1in]{geometry}
\usepackage{ARAstroBib,amsmath,amssymb,graphicx,deluxetable}

\begin{document}
\input epsf.tex
\input epsf.def
\input psfig.sty

\jname{}
\jyear{}
\jvol{}
\ARinfo{}

\title{Thermonuclear burst oscillations}
\markboth{Anna L. Watts}{Thermonuclear burst oscillations}

\author{Anna L. Watts
\affiliation{Astronomical Institute ``Anton Pannekoek'', University of Amsterdam, \\
Science Park 904, 1090 GE Amsterdam, the Netherlands}}


\begin{abstract}
Burst oscillations, a phenomenon observed in a significant fraction of Type I (thermonuclear) X-ray bursts, involve the development of highly asymmetric brightness patches in the burning surface layers of accreting neutron stars.    Intrinsically interesting as nuclear phenomena, they are
also important as probes of dense matter physics and the strong gravity, high magnetic
field environment of the neutron star surface.   Burst oscillation frequency is also used to measure stellar spin, and doubles the sample of rapidly rotating (above 10 Hz) accreting neutron stars with known spins.  Although the mechanism remains mysterious, burst oscillation models must take into account thermonuclear flame spread, nuclear processes, rapid rotation, and the dynamical role of the magnetic field.  This review provides a comprehensive summary of the observational properties of burst oscillations, an assessment of the status of the theoretical models that are being developed to explain them, and an overview of how they can be used to constrain neutron star properties such as spin, mass and radius.
\end{abstract}

\maketitle

\section{Introduction}
\label{intro}

Type I X-ray bursts are thermonuclear explosions triggered by unstable
burning in the oceans of accreting neutron stars.  The ocean is the low density fluid layer of
  light elements that builds up via accretion on top of the solid
   crust (see \citealt{Chamel08} for a discussion of the
  conditions under which a Coulomb plasma of ions will crystallize to
  form a solid).  The basic cause of bursts, an imbalance
between nuclear heating and radiative cooling in settling material, can be explained using
simple one-dimensional calculations (for reviews see \citealt{Lewin93}
and \citealt{Bildsten98b}).   The past decade, however, has revealed major
complexities that can no longer be understood within the context of
such simple models \citep{Strohmayer06}.  

One such complexity is the development of asymmetric brightness patches, known as burst oscillations, in about 10\% of the bursts observed with high time resolution instrumentation \citep{Galloway08}.  What drives this process remains unexplained, and requires us to consider flame spreading and other multidimensional effects.  The highly atypical properties of burst oscillations from the accretion-powered pulsars also suggest an important dynamical role for the magnetic field.  This review article provides an overview of our current understanding
of burst oscillations:  the techniques employed to detect and analyse
them, their key observational properties, the status of theoretical
models, and the ways in which they can be used to constrain stellar spin rates and the dense
matter equation of state.     

\subsection{Burst oscillations - a brief history}

Although a number of searches for periodic phenomena in X-ray bursts were carried out prior to the mid 1990s, and some tentative detections were claimed, none have stood the test of time (\citealt{Lewin93}, \citealt{Jongert96}).  Conclusive
detection of strong periodic signals in thermonuclear bursts came only with
the launch of the {\it Rossi X-ray Timing Explorer} (RXTE) on December
30th, 1995.  

Observations of the burster 4U 1728-34 in February 1996, only weeks
after the start of science operations, led to the discovery of a
strong 363 Hz signal in six X-ray bursts
\citep{Strohmayer96b}. The
authors dubbed this phenomenon `burst oscillations'.  Many
of the properties that we now consider hallmarks of burst
oscillations were apparent even in these first observations: high
coherence, an upwards 
drift of $\sim 1$ Hz towards an asymptotic maximum frequency as the burst progressed, amplitudes of up to $\sim 10$
\% root mean square (rms), and the 
apparent disappearance of the signal during the burst peak.  
  
\citet{Strohmayer96b} concluded that rotational modulation of a bright
spot on the burning surface was the most likely explanation for the
burst oscillations.  This was based on three arguments: the
expected evolutionary link between accreting neutron stars in Low Mass X-ray Binaries and the
millisecond radio pulsars (with the former being progenitors of the latter
via the spin recycling scenario, see \citealt{Bhattacharya91}); the
ability of a localised hotspot to explain the observed amplitudes; and
the high coherence of the oscillations seen in the burst tails. 

Between 1996 and 2002, burst oscillations were discovered in
bursts from eight more sources (see Table \ref{sources}).  Although there
was still no independent confirmation of spin rate, the link between
burst oscillation frequency and
stellar rotation was nonetheless strengthened. The fact that
the same frequency was seen in multiple bursts from any given source,
and the high stability of the asymptotic frequency of the
drifting burst oscillations, both pointed to
rotational modulation of a bright spot that was near stationary in the
rotating frame of the star (\citealt{Strohmayer98b}, \citealt{Muno02a}).  The
detection of burst oscillations during a superburst
(a longer, more energetic burst thought to be due to unstable
carbon burning) from 4U 1636-536 also supported this interpretation.  The frequency matched that seen in the normal Type I bursts despite
the different burst physics, suggesting an external clock
\citep{Strohmayer02a}. 

In October 2002, the 401 Hz accretion-powered millsecond X-ray pulsar SAX
J1808.4-3658 went into outburst, and \citet{Chakrabarty03} reported the first robust detection of burst oscillations from a source with an independent measure of the
spin rate (analysis of a burst during a previous accretion episode had yielded a marginal detection, \citealt{intzand01}).  Although the frequency of the burst
oscillations drifted by several Hz in the burst rise, the frequency in
the tail was within $\approx 6 \times 10^{-3} $ Hz of the spin
frequency.  A second accretion-powered pulsar with burst
oscillations at the spin frequency, XTE J1814-338, followed shortly thereafter
\citep{Strohmayer03}, confirming the status of burst oscillation
sources as nuclear-powered pulsars.  

Since this time, burst oscillations have been found in several
additional sources (Table \ref{sources}), including three more
accretion-powered pulsars and two intermittent pulsars (sources
that show accretion-powered pulsations only sporadically). While
rotational modulation of a brightness asymmetry on the burning surface remains an integral
part of all models, the root cause of such an asymmetry remains
unresolved.

\section{Burst oscillation observations}

\subsection{Principles of detection}
\label{detection}

The standard analysis technique, when searching for periodic or
quasi-periodic signals, is to use Fast Fourier Transforms to produce a
power spectrum.  For a comprehensive review of this topic, the reader
is referred to \citet{vanderKlis89}.  Here I summarize only the key
points that are essential to an understanding of burst oscillation
detection (this is of particular relevance to the discussion in
Table \ref{tentsources} about tentative detections) and subsequent analysis of
properties like frequency and amplitude.  

A series of X-ray photon arrival times with total duration $T$ is binned 
to form a time series $x_k(t)$, the number of photons (`counts') in time bin
$t_k$ ($k = 1...N$).  The time resolution $\Delta t = T/N$ is limited
by the native time resolution of the instrument 
(typically 1 - 125 $\mu$s for RXTE, depending on data mode), but is
also set by the desired Nyquist frequency $f_\mathrm{Ny}$.  This is the maximum frequency that can be studied for a given time resolution, and is given by $f_\mathrm{Ny} = 1/(2\Delta t)$.   The
power spectrum $P_j$ at the Fourier frequencies $\nu_j = j/T$ ($j = 0,
2, ..., N/2$ where $\nu_{N/2} = f_\mathrm{Ny}$), using the standard Leahy
normalisation \citep{Leahy83}, is then given by 

\begin{equation}
P_j = \frac{2}{N_\gamma} \left[\left(\sum^N_{k=1} x_k \cos
    2\pi \nu_j t_k\right)^2 + \left(\sum^N_{k=1} x_k \sin 2\pi
    \nu_jt_k\right)^2\right]
\end{equation}
where $N_\gamma$ is the total number of photons.  

In the absence of any deterministic signal, the Poisson statistics of
photon counting yield powers that are distributed as $\chi^2$
with two degrees of freedom (d.o.f.).  The presence of a periodic signal at a given frequency will boost the power in the appropriate frequency bin.  If we detect a large power, however, we must first evaluate its significance by computing the probability of obtaining such a power through noise alone. By this we mean the
chances of getting such a high value of the power, 
in the absence of a periodic signal, due to the natural fluctuations
in powers that are a by-product of photon counting statistics.  We
compute this using the known properties of the $\chi^2$ distribution,
taking into account the numbers of trials (see below).

Many burst oscillation papers use the $Z_n^2$ statistic, instead of the
power spectrum.   Although very similar to the standard power spectrum computed from a Fourier
transform, it does not require that the photon arrival times be binned.  It is
defined as

\begin{equation}
Z^2_n = \frac{2}{N_\gamma}\sum^n_{k=1}\left[\left(\sum^{N_\gamma}_{j=1} \cos
    k\phi_j\right)^2 + \left(\sum^{N_\gamma}_{j=1} \sin
    k\phi_j\right)^2\right]
\end{equation}
where here $n$ is the number of harmonics (most commonly taken to be 1), $k = 1, 2, ... n$ is the index used to sum over harmonics,  $N_\gamma$ is the total number of photons, and $j$ is an index applied to
each photon.  The phase $\phi_j$ for each photon is defined as

\begin{equation}
\phi_j = 2\pi \int_{t_0}^{t_j} \nu(t) dt
\label{phasemodel}
\end{equation}
where $\nu(t)$ is the frequency under consideration and
$t_j$ is the arrival time of each photon relative to some reference
time $t_0$.   Frequency $\nu(t)$ is most commonly taken to be constant (so that it can be taken outside the integral) but  on occasion may be a function of time, for example if one wants to correct for smearing due to orbital Doppler shifts as the neutron star orbits the center of mass of the binary.  In the absence of a deterministic (e.g. periodic) signal, $Z_n^2$ is distributed as
$\chi^2$ with $2n$ d.o.f..   An advantage of $Z_n^2$ is that it offers an efficient way of summing harmonics.  

When searching for drifting, or quasi-periodic signals, it is
common to average powers from neighbouring frequency bins.   In searches for weak
signals one can also stack, or take an average of, power
spectra from many independent data segments (time windows) or bursts.  This affects
the number of degrees of freedom in 
the theoretical $\chi^2$ distibution of noise powers, but the modified
theoretical distribution is known (see
\citealt{vanderKlis89} for more details).  

To assess whether a periodic signal is indeed present, we first take the power spectrum and identify the frequency bins with the highest powers.  One then
 computes the probability of obtaining such high powers
through Poisson noise alone, using the properties of the $\chi^2$ distribution
with the appropriate number of degrees of freedom.  One must then take
into account the number of trials that have been made - the more trials, the more likely one is to obtain a high power via statistical fluctuations alone.  The number of trials is the product of the number of 
independent frequency bins, time windows, energy bands, bursts and (if appropriate) sources searched.  Assessing the number of trials
properly is complicated by the fact that many burst oscillation
searches use overlapping time bins or oversampled frequency bins to
maximise candidate signal power.  Failure to account properly for this
can result in the significance of a candidate signal being
overestimated (or underestimated, if overlapping bins are treated as being independent).  If the probability of such a high value of the power arising through
Poisson noise alone is below a certain threshold after taking into account
numbers of trials, then it is deemed significant, with the quoted
significance referring to the chances of having obtained such a signal
through Poisson noise alone.   

A complication comes from the fact that noise powers (the power distribution in the absence of a periodic signal) are
not always distributed in accordance with the $\chi^2$ that would be expected from constant Poisson noise.  What we are
searching for in an X-ray burst is a periodic signal  superimposed on
the overall deterministic rise and decay of a burst lightcurve.    These variations give low frequency power and add sidebands to noise powers, boosting their level (for a nice discussion of these issues, see
\citealt{Fox01}).   This problem is particularly acute in the rising 
phase of the burst when the lightcurve is changing rapidly. There may also be other noise components from astrophysical or detector processes, such as red noise associated with accretion, which may continue during the burst and contribute to the overall emission.

Assessing the true distribution that the powers should take, in the absence of a periodic signal, is
most easily done using Monte Carlo simulations.  One can for example make a model
of the burst lightcurve, by fitting the rise
and decay with exponential functions, and then use this as a basis to generate a large sample of
fake lightcurves with Poisson counting statistics but without any periodic signals.
Taking power spectra from these fake lightcurves then yields the true, most likely
frequency-dependent, distribution of noise powers.  One other
advantage of the Monte Carlo method is 
that any pecularities of the data analysis process (such as the use of
overlapping time windows) can be replicated precisely.  As a rule of
thumb, unless a
candidate signal is
particularly strong, or is repeated in multiple independent
bursts or time windows (thereby boosting
significance), Monte
Carlo simulations should really be done to obtain an accurate
assessment of significance.   Alternatively, one can also try to fit the distribution of measured powers (excluding the frequency bins containing the candidate signals), however this is somewhat risky since it by definition excludes extreme values from the fit. 

One final factor that must be considered, for the accretion-powered pulsars
with burst oscillations, is the likely continuation of the channeled
accretion process during the burst.   One must therefore ask
whether the signal detected during the burst could be due to the
accretion-powered pulsations.   If the accretion process (accretion rate, and fractional amplitude of accretion-powered pulsations) remains unchanged during the burst, the addition of a large number of unpulsed photons during the burst would cause the measured fractional amplitude to fall (by a factor $N_\mathrm{bur}/N_\mathrm{acc}$ where $N_\mathrm{bur}$ is the number of photons due to burst emission and $N_\mathrm{acc}$ the number due to accretion in the data segment being considered).  In all cases so far, fractional amplitudes remain similar despite the fact that burst
flux far exceeds accretion flux, indicating that the burst emission itself
must also be pulsed (provided that the assumption of unchanged
accretion holds).    Whether this assumption is reasonable remains a matter of debate. Naively one might expect radiation 
pressure from a burst to hinder the accretion process. However the radiation may also 
remove angular momentum, increasing the accretion rate (\citealt{Walker89},  \citealt{Miller93}, \citealt{Ballantyne05}). 

\subsection{Measurable properties}

Once a burst oscillation signal has been identified and deemed statistically
significant, one can measure several observational properties.    Some
properties can be calculated directly from the power spectrum or $Z_n^2$ statistic (for the rest of this subsection, these two
  terms can be used interchangeably), whereas
others require pulse profile modelling.  The latter technique involves selecting a
frequency model for the signal, $\nu(t)$, and calculating a phase for each photon using Equation (\ref{phasemodel}), using this to assign
each photon to a phase bin, and
thereby building up a pulse profile (number of photons per phase bin).    The main intrinsic properties of interest are frequency,
amplitude, and phase lags.  The first two can be measured from the
power spectrum or using pulse profile modelling, but to obtain phase lags one must use pulse
profile modelling since the power spectrum discards phase information.  

\begin{figure}
\begin{center}
\includegraphics[width=\columnwidth, clip]{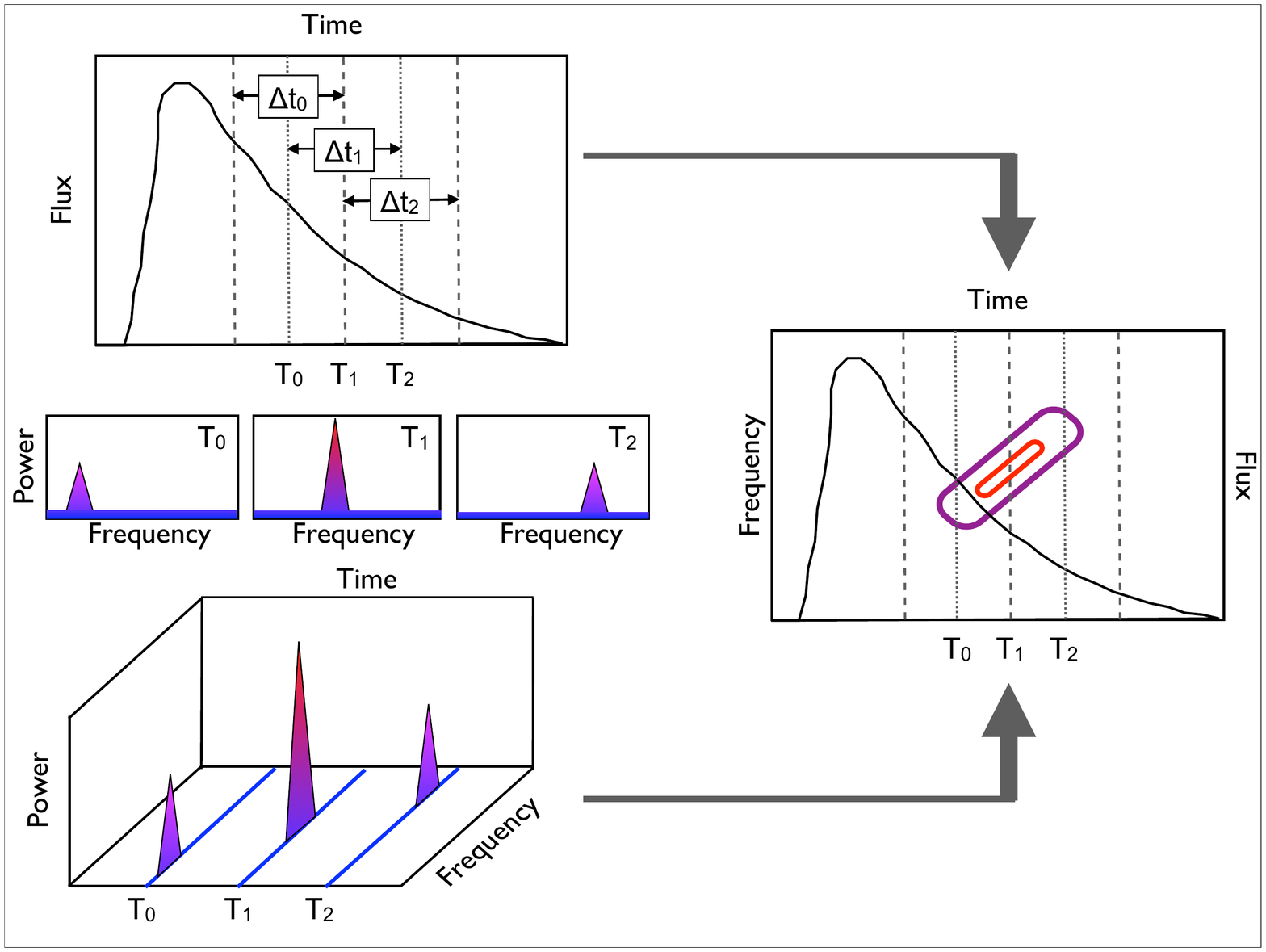}
\caption{Generating a dynamical power spectrum.  The burst lightcurve (solid black line) is first split into segments (top left).  In this case, there are three overlapping time windows, $\Delta t_0$, $\Delta t_1$ and $\Delta t_2$.  The time stamps $T_0$, $T_1$ and $T_2$ refer to the midpoints of each window.  Computing a power spectrum for each segment (center left), one can see that both power and frequency vary with time (blue = low power, purple = medium power, red = high power).  This can be plotted in power-frequency-time space (bottom left).  The resulting contours of power (purple = medium, red = high) are projected onto the frequency-time plane and the resulting dynamical power spectrum overplotted on the lightcurve (right). }
\label{dps}
\end{center}
\end{figure}

\begin{figure}
\begin{center}
\includegraphics[width=\columnwidth, clip]{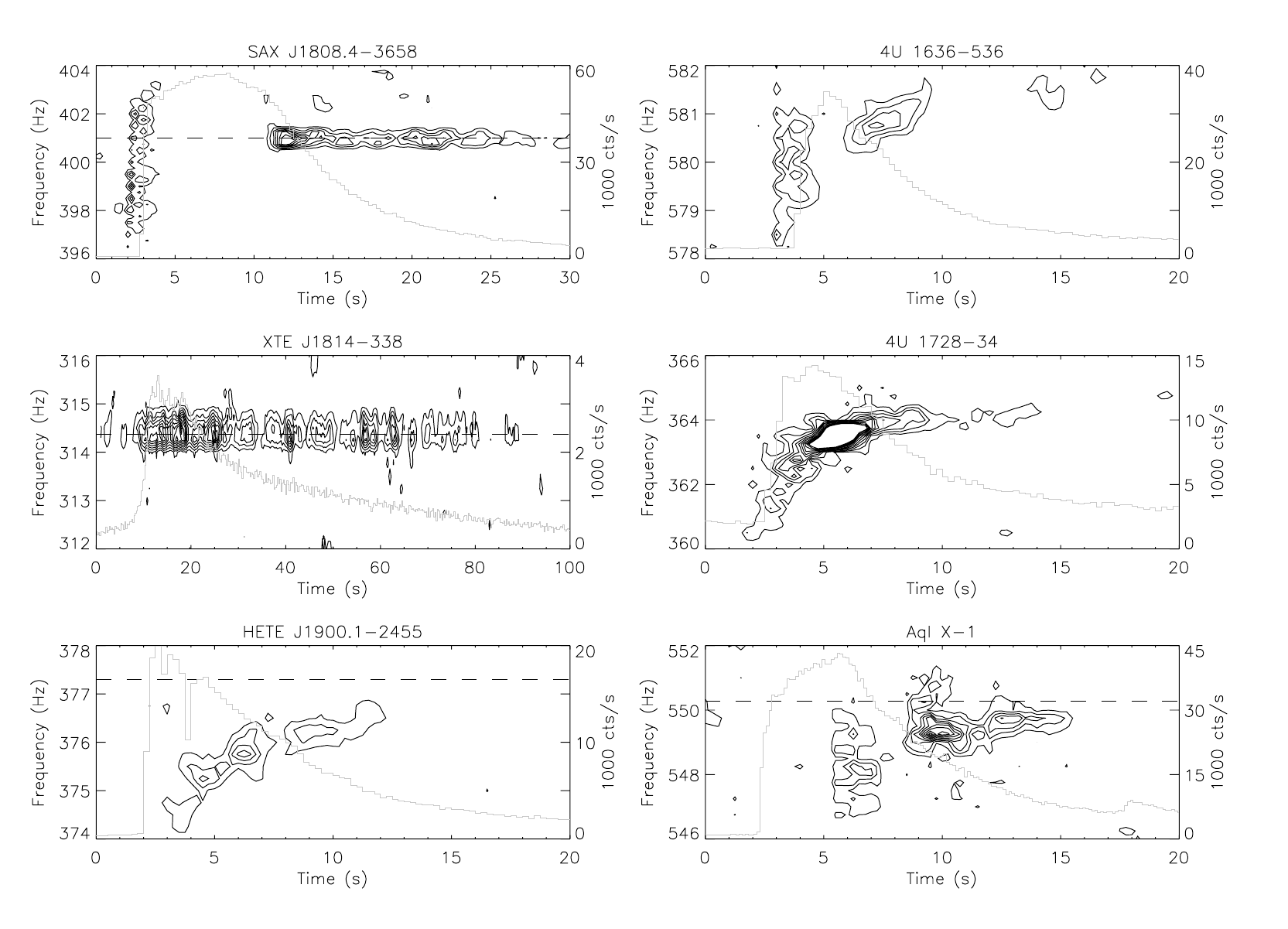}
\caption{Light curves (grey) and dynamical power spectra (black) showing burst oscillations for bursts from several sources:  two persistent accretion-powered pulsars (SAX J1808.4-3658 and XTE J1814-338), two intermittent pulsars (HETE J1900.1-2455 and Aql X-1) and two non-pulsars (4U 1636-536 and 4U 1728-34), using data from RXTE.  The dynamical power spectra use overlapping 2 s windows, with new windows starting at 0.25 s intervals.  We use a Nyquist frequency of 2048 Hz and an interbin response function to reduce artificial drops in amplitude as the frequency drifts between Fourier bins.  The contours show Leahy normalized powers of 10-100, increasing in steps of 10.   The dashed lines on the pulsar plots indicate the spin frequency determined from accretion-powered pulsations. }
\label{bplots}
\end{center}
\end{figure}

In terms of frequency, we are interested in both the absolute value
and the properties of any drift.  The latter is most commonly
visualised using dynamical power spectra.  These are computed by taking
power spectra of short, usually overlapping, segments of data from the
burst lightcurve, as illustrated in Figure \ref{dps}.  The power spectra
 are then commonly plotted as contours, overlaid on the
burst lightcurve.  Several examples of bursts with oscillations, from different sources, are shown Figure \ref{bplots}.  One can then see the
frequency drift that occurs as the burst progresses.  Another
property, related to frequency, that is often given is the coherence
$Q$ 
of the burst oscillation train.  This is given by

\begin{equation}
Q = \frac{\nu}{\Delta \nu_\mathrm{FWHM}},
\end{equation}
where $\nu$ is the frequency of the peak in the power spectrum and
$\Delta \nu_\mathrm{FWHM}$ its full width at half-maximum, obtained by
fitting a Gaussian or Lorentzian function to the broad peak in the
power spectrum.  Note that $Q$ values are often quoted for peaks where the frequency drift has
been incorporated into the model (as in Equation \ref{phasemodel}), thereby rendering the peak in the power
spectrum narrower, and hence more coherent (see for example
\citealt{Strohmayer99}).  In this case the frequency $\nu$ that is
used in computing $Q$ is normally the
asymptotic maximum frequency (Sections \ref{intro} and \ref{frequencies}).  

The amplitude of the pulsations can be computed directly from the power
spectrum, or from a folded pulse profile (see for example \citealt{Muno02b} and \citealt{Watts05}).  When using the power spectrum, the root mean square (rms)
fractional amplitude $r$ is given by

\begin{equation}
r = \left(\frac{P}{N_\gamma}\right)^\frac{1}{2}
  \left(\frac{N_\gamma}{N_\gamma - N_b}\right)
\end{equation}
where $P$ is the signal power, $N_\gamma$ the total number
of photons, and $N_b$ the number of background photons (estimated from the period before or after the burst).  Note that signal power is not
  the same as the measured power.  Once a power has been deemed
  significant (i.e. unlikely to have occurred through noise alone),
  one must then determine the signal power that would give rise to the
  measured value, since measured power includes the effects of noise
  \citep{Groth75}.   For details of how to compute signal power from measured power, see \citet{Vaughan94}, \citet{Muno02b} or \citet{Watts05}.  For sources
that also have accretion-powered pulsations, this value is often
corrected to remove the expected contribution due to 
continuing pulsed accretion during the burst.  Other related
quantities that can be computed are the amplitudes of any higher
harmonics of the burst oscillation frequency (often referred to as the
`harmonic content'), and the dependence of
the amplitude on photon energy.  

There are several factors to be wary of when using values of amplitude
from the literature.  Different groups use different definitions of amplitude:  for a lightcurve that varies as $C + A \sin  2\pi \nu t$, different groups may quote, for the signal at frequency $\nu$, full fractional amplitude ($2A/C$), half fractional amplitude ($A/C$), or root mean square (rms) fractional amplitude ($A/C\sqrt{2}$).  In addition, amplitude depends on the frequency model used.  Fitting apparent
frequency drift will increase amplitude, but this must be taken into
account when testing models since the two are not independent.  The
third factor, specific to the power 
spectrum technique, is that amplitude is suppressed when frequency
drifts between Fourier bins \citep{vanderKlis89}.  This effect can
sometimes be seen in published dynamical power spectra:  compare for example the
dynamical power spectra in Figure 1 of \citet{Muno02b} with those in
Figure 1 of \citet{Watts09}, where an
interbin response function is used to reduce the effects of this
drop (although note that this does not affect the amplitudes in \citet{Muno02b}, which
  were computed using folded pulse profiles).

Energy-dependent phase lags, which are of use when testing models, are computed by
comparing pulse profiles constructed from photons in different energy
bands.  The degree to which the folded pulse profiles are offset is the phase lag.
This type of analysis can also be used to compare the phases of burst
oscillations to accretion-powered pulsations, for sources that show
both phenomena.   Care must be taken when doing this to correct for
any contamination due to the continued presence of accretion-powered
pulsations during the burst.  However the magnitude of the correction
is simple to estimate (see \citealt{Watts08b}). 

Finally, one can analyse the dependence of the burst oscillation
characteristics on the properties of
the bursts and the accretion state of the star.  Burst
properties may include the total integrated flux (fluence), peak flux, the presence or absence
of photospheric radius 
expansion, duration, and rise and decay timescales.  In terms of accretion
state, one can consider the overall accretion rate (which can be estimated in various ways) and, for the
pulsars, compare the properties of the burst oscillations to those of
the accretion-powered pulsations.  Note that burst type also depends
strongly on accretion rate, so these two factors are not unrelated.  

\subsection{Observational properties}

\subsubsection{Overview of burst oscillation sources}

Secure detections of burst oscillations have now been made in 17
sources (Table \ref{sources}), including 5 persistent
accretion-powered pulsars and 2 intermittent accretion-powered
pulsars.   Note that all are transient accretors, and the use of the term persistent to describe the presence of accretion-powered pulsations
  through accretion episodes - as used here - 
  should not be confused with its use to describe sources that are
  persistently, as opposed to transiently, accreting.  All of the burst oscillation sources are in Low Mass X-ray Binaries \citep{Tauris06} with orbital periods (where known) in the range 1.4 to 19 hours.  There is one candidate ultracompact binary in this group, 4U 1738-34, but the orbital period for this source is not known, see \citealt{Galloway10b}.  Ideally, for a detection to be secure, the same frequency
should have been seen in multiple bursts from the same source.  For
one of the sources in Table \ref{sources} this is not the case.  In
this case the high statistical significance of the detection is based
upon the fact 
that the same frequency was seen in multiple independent time bins
within a single burst, very close to the spin frequency
observed from accretion-powered pulsations.     

Table \ref{tentsources} gives details of several other sources for
which burst oscillation detections 
have been claimed.  Some of the detections are very marginal (below
the standard 3$\sigma$ threshold), and most
rely on a power in a
single independent time bin in a single burst.  Also given in the
table is a summary of the analysis procedure used to estimate the
significance of the claimed detection.  Comparing these procedures to
the ideal outlined in Section \ref{detection},
the significance of some of these results has probably been
over-estimated. As such, they should be considered tentative until
confirmed in a second burst.   
 
\begin{deluxetable}{l c p{10cm}}
\tabletypesize{\scriptsize}
\tablewidth{0pt}
\tablecaption{Sources with confirmed detections of burst oscillations \label{sources}}
\tablehead{\colhead{Source}  & \colhead{Frequency (Hz)} &
  \colhead{References\tablenotemark{\dag}} \\}
\tablecolumns{3}
\startdata
\sidehead{Persistent accretion-powered pulsars with burst
  oscillations}
SAX J1808.4-3658 & 401 & \citet{Chakrabarty03}  \\
IGR J17498-2921 & 401 & \citet{Linares11}, \citet{Chakraborty12} \\
XTE J1814-338 & 314 & \citet{Strohmayer03}  \\
IGR J17511-3057 & 245 & \citet{Altamirano10a} \\
IGR J17480-2446 & \phantom{ }11 & \citet{Cavecchi11} \\
\sidehead{Intermittent accretion-powered pulsars with burst oscillations}
Aql X-1 & 550 & \citet{Zhang98}  \\
HETE J1900.1-2455 & 377 & \citet{Watts09}\tablenotemark{\ddag}   \\
\sidehead{Burst oscillation sources without detectable accretion-powered pulsations}
4U 1608-522 & 620 & \citet{Hartman03}, \citet{Galloway08}  \\
SAX J1750.8-2900 & 601 & \citet{Kaaret02}, \citet{Galloway08} \\
GRS 1741.9-2853 & 589 &  \citet{Strohmayer97b} \\
4U 1636-536 & 581 & \citet{Strohmayer98a}, \citet{Strohmayer02a} \\
X 1658-298 & 567 & \citet{Wijnands01} \\
EXO 0748-676 & 552 & \citet{Galloway10} \\
KS 1731-260 & 524 & \citet{Smith97}, \citet{Muno00}  \\
4U 1728-34 & 363 & \citet{Strohmayer96b} \\
4U 1702-429 & 329 & \citet{Markwardt99} \\
IGR J17191-2821 & 294 & \citet{Altamirano10b} \\
\enddata
\tablenotetext{\dag}{Second (confirmation) references are given for
  sources where the initial discovery rested on a single burst (but
  see also note \ddag).}
\tablenotetext{\ddag}{Although burst oscillations from this source have only
  been detected in a single burst, they were observed in multiple
  independent time bins.  We
  therefore consider this detection to be secure.}
\end{deluxetable}

\begin{deluxetable}{l c p{11cm}}
\tabletypesize{\tiny}
\tablewidth{0pt}
\tablecaption{Sources with single burst or tentative burst
  oscillation detections  \label{tentsources}}
\tablehead{\colhead{Source}  & \colhead{Frequency (Hz)} & \colhead{Remarks}  \\ }
\tablecolumns{3}
\startdata
XTE J1739-285 & 1122  & \citet{Kaaret07} report a $\sim 4 \sigma$
signal in a 4s window in the tail of one burst 
(of five observed). The significance was estimated using Monte Carlo
simulations, taking into account number of bursts, energy
bands, frequencies and time windows 
searched. \citet{Galloway08}, however, report no significant signal in
the same data. The result seems very sensitive to the choice of
time windows (start point, overlapping vs. independent).   \\ 

GS 1826-34 & \phantom{ }611 & \citet{Thompson05} report a $\sim 4\sigma$ signal after stacking power spectra from multiple 0.25 s time windows from 3 separate bursts, fitting the measured powers to obtain the distribution and taking into account the number of bursts, time windows and frequency bins used to create the power spectra.   However the number of trials is underestimated:  the authors made selections in energy, in addition to searching other time windows and fine-tuning both segment length and the number of segments included.  \\

A 1744-361 & \phantom{ }530  & \citet{Bhattacharyya06d}
report a $\sim 6\sigma$ signal in a 4s window in the
rise of the one burst observed from this source. This
significance is estimated using $\chi^2$ with 2 d.o.f., taking into account the number of frequencies searched. \citet{Galloway08} confirm the detection in this burst,
but no other burst from this source has been observed with high time
resolution instruments.\\

4U 0614+09 & \phantom{ }415 & \citet{Strohmayer08} report
a $\sim 4 \sigma$ signal in a 5s window in the cooling tail of one
burst (of two observed). This significance is estimated using $\chi^2$ with 2 d.o.f., taking into account frequency and
time windows searched.  \\

SAX J1748.9-2021 & \phantom{ }410 & \citet{Kaaret03} report a $4.4 \sigma$ oscillation in one time window of one burst from this source.  Significance was estimated using the $\chi^2$ distribution with 2 d.o.f., taking into account the number of frequency bins searched.   This source has since been discovered to be an intermittent accretion-powered pulsar with spin frequency 442 Hz \citep{Altamirano08}.  Re-analysis by \citet{Altamirano08}, including the number of bursts and time windows searched in the number of trials, and not using overlapping windows to maximise power, found a revised significance for the burst oscillation candidate of $2.5\sigma$. \\

MXB 1730-335 & \phantom{ }306  & \citet{Fox01} report a
$\sim 2.5\sigma$ candidate signal after stacking power
spectra from
the 1s rising phase of 31 bursts.  This significance is estimated
using
Monte Carlo simulations of the stacked spectra, taking into account
the number of frequency
bins searched.\\

4U 1916-053 & \phantom{ }270 & \citet{Galloway01} report a $4.6\sigma$ signal
during the $\sim 1$s rise of one burst. Significance is estimated using $\chi^2$ with 2 d.o.f., taking into account the number of frequency bins
searched.  Weaker candidate signals are
present in adjacent frequency bins in the tail, but the signal cannot
be phase-connected between rise and tail.  Burst oscillations are not 
seen in the 13 other bursts from this source observed by
RXTE \citep{Galloway08}. \\

XB 1254-690 & \phantom{  }95 & \citet{Bhattacharyya07b} reports a $2\sigma$
candidate signal in the 1s rise of one burst (of five observed).  This
significance is estimated using $\chi^2$ with 2 d.o.f., taking into account bursts and frequency bins
searched. \\

EXO 0748-676 & \phantom{  }45 & \citet{Villarreal04} report a $\sim 5.5\sigma$ signal after stacking power spectra from the tails of 38 bursts.  Significance was estimated by fitting the distribution of measured powers, taking into account the number of frequency bins searched.  Since this time, strong burst oscillations at 552 Hz have been discovered in this source (\citealt{Galloway10} and Table \ref{sources}).  \citet{Galloway10} confirm the detection of the 45 Hz feature in the sample used by \citet{Villarreal04}, but do not find it when using a larger sample of bursts.  Its nature remains unclear. 

\enddata
\tablecomments{The table gives the significance quoted by the authors
  for the candidate detection, and explains the methods used to derive
  this significance. As explained in Section \ref{detection}, estimates
  derived using $\chi^2$ with 2 d.o.f. rather than Monte
  Carlo simulations tend to overestimate significance.  The procedure
  for estimating number of trials varies from paper to paper: ideally,
  proper allowance should be made for the number of bursts,
  energy bands, frequency bins {\bf and} time windows searched.   The SAX J1748.9-2021 result is particularly informative in terms of illustrating the importance of using the correct number of trials.   Whether to
  give equal weight in terms of numbers of trials to every burst
  searched remains a matter of debate.  The number of RXTE proportional counter units active
  during observations varies (with more photons maximising detection
  chances), and in addition burst oscillations tend to occur
  preferentially in certain accretion states (Section \ref{burstcond}). } 

\end{deluxetable}

\subsubsection{Conditions in which burst oscillations are detected}
\label{burstcond}

For four of the five persistent pulsars, burst oscillations have been detected in
all observed bursts, irrespective of burst properties or accretion
state (see Table \ref{sources} for references) - although the persistent pulsars do not typically exhibit a wide range of accretion states.  For the fifth persistent pulsar (IGR J17498-2921), oscillations have only been detected in the two brightest bursts.  However the upper limits on the presence of oscillations in the weaker bursts are comparable to the amplitudes detected in the brighter bursts \citep{Chakraborty12}, so they could well be present at the same level in these other bursts.  For the other sources (the non-pulsars and intermittent pulsars), burst oscillations are detected in only a subset of bursts.  

A comprehensive analysis of burst and burst oscillation 
properties by \citet{Galloway08} found that the properties of the
bursts (e.g. duration, recurrence 
times) where oscillations are detected are not
unusual compared to the properties of bursts where oscillations
are not detected.  There does however appear to be a correlation with
accretion state, as estimated from the color-color diagram, a plot of hard against soft X-ray colors.  As accretion rate increases, Low Mass X-ray Binaries typically move from the top left (hard) to bottom right (soft), tracing out a Z-shaped pattern \citep{Hasinger89}.  The persistent pulsars, including those with burst oscillations, tend to remain in the hard (low accretion rate) state.  The sources that are not persistent pulsars tend to show burst oscillations when the source is in the soft (high accretion rate) state.

This leads, as previously noted by \citet{Muno01} and \citet{Muno04}, to
an apparent relationship between the occurrence of burst oscillations, spin frequency, burst type
and the presence of photospheric radius expansion (PRE).  Sources with burst
oscillation frequency $< 400$ Hz tend to have short, most likely helium-dominated
bursts.  Higher frequency sources tend to have longer bursts, that probably involve mixed
hydrogen/helium burning.    In the soft state, short bursts are less likely to have PRE whilst long bursts are more likely.   Since burst oscillations occur more often in the soft state, the low frequency sources are more likely to have oscillations in bursts without PRE, while high frequency sources are more likely to
have oscillations in bursts with PRE.  The distinction is however not absolute and the relationship between these various factors is clearly complex.  For a more in-depth discussion of the apparent link between burst type and rotation rate, see \citet{Galloway08}.   

Given that burst oscillations (for sources that are not
persistent pulsars) seem to occur preferentially in certain accretion states, one
can ask whether the non-detection of burst oscillations in other
sources is at all surprising.   \citet{Galloway08} characterized position on the Z-shaped track in the color-color diagram using a variable $S_z$, with low values of $S_z$ corresponding to harder states, and high
values to softer states.  For 87\% of bursts with oscillations, they found $S_z>1.75$, cementing the link between burst oscillations and the soft, high accretion rate state.   Table \ref{otherbursters} summarizes 
the status for prolific bursters in their data set, for which burst oscillations have not been found.  The
sample includes one intermittent pulsar, SAX J1748.9-2021.  For most of
these sources the range of accretion states that were observed was
not sufficient to enable full characterization of the color-color
diagram.  This means that the position variable $S_z$ could not be calculated for these
sources, and so it is difficult to judge whether the non-detection of burst oscillations
is unexpected.  For the two sources that do have a
fully-characterized color-color diagram (4U 1705-44 and 4U 1746-37), however, there are bursts with $S_z>1.75$.  Given that 57\% of the bursts with $S_z>1.75$ studied by \citet{Galloway08} showed burst
oscillations, it is perhaps rather surprising that no
oscillations have been found in these two sources (although the source with the most high $S_z$ bursts, 4U 1746-37, has very
unusual bursting behaviour in all respects).   A more rigorous analysis of whether the non-detection of oscillations in the other bursters can be attributed to their being observed in harder accretion states would be desirable.   

In terms of when burst oscillations occur during bursts, \citet{Galloway08} show
that although they can be observed at any point during the burst
(rise, peak or tail), they are most commonly detected in
the tails of bursts.   Oscillation trains tend to be interrupted during the
peaks of bursts, particularly (although by no means exclusively) during episodes of
PRE (\citealt{Galloway08}, \citealt{Altamirano10a}).  

Burst oscillations are clearly a common feature of many normal Type I X-ray
bursts.   But what about the other types of thermonuclear burst?  A number of systems (although none
of the confirmed burst oscillation sources in Table \ref{sources})
have shown intermediate 
duration bursts that last several hundred seconds (\citealt{intZand07},
  \citealt{Falanga08}, \citealt{Linares09}, \citealt{Kuulkers10}).  At present there are two models to explain
the occurrence of such bursts, both involving the build-up and 
ignition of a thick layer of helium.  Either the system is
ultracompact, so that it accretes nearly pure helium (\citealt{intZand05}, \citealt{Cumming06})
or a thick layer of helium is built up from unstable hydrogen burning
at low accretion rates (\citealt{Peng07}, \citealt{Cooper07c}).  Due to their
rarity, very few intermediate bursts have been observed with high time
resolution instruments.  Where they have, no significant burst
oscillation signal has been found (\citealt{Linares09}, \citealt{Kuulkers10}).  

Burst oscillations have however been detected in one superburst
\citep{Strohmayer02a}. The source in question, 4U 1636-536, also shows burst oscillations at the same
frequency in its Type I bursts.  Superbursts, which last several
hours, are thought to be caused by unstable carbon ignition deeper
within the ocean than the Type I burst ignition point
(\citealt{Cumming01}, \citealt{Strohmayer02b}, \citealt{Kuulkers04}).  The oscillations were
detected over about 800 s during the peak of the 4U 1636-536
superburst, but not during the long, decaying tail.  High time resolution
data has only been obtained for one other superburst, from the
ultracompact source 4U 1820-30.  Unlike 4U 1636-536, this source has
not shown burst oscillations in its Type I bursts.  Unfortunately an antenna
malfunction on {\it RXTE} resulted in the loss of the high time
resolution data from the peak of this superburst \citep{Strohmayer02b}.  However no oscillations
are detected during the superburst tail (unpublished analysis).  

\begin{deluxetable}{lccp{8cm}}
\tabletypesize{\scriptsize}
\tablewidth{0pt}
\tablecaption{Prolific bursters without
  oscillations. \label{otherbursters}} 
\tablehead{\colhead{Source}  & \colhead{Number of bursts} &  $S_z$ reported? &  Bursts
  with $S_z>1.75$  }
\tablecolumns{4}
\startdata
2E~1742.9$-$2929   & 84 & N & \nodata \\
Rapid~Burster      & 66 & N & \nodata \\
GS~1826$-$24       & 65 & N & \nodata \\
Cyg~X-2            & 55 & N & \nodata \\
4U~1705$-$44       & 47 & Y & 8 \\
4U~1746$-$37       & 30 & Y & 20 \\
4U~1323$-$62       & 30 & N & \nodata  \\
SAX~J1747.0$-$2853 & 23 & N & \nodata  \\
EXO~1745$-$248     & 22 & N & \nodata  \\
XTE~J1710$-$281    & 18 & N & \nodata  \\
1M~0836$-$425      & 17 & N & \nodata \\
SAX~J1748.9$-$2021 & 16 & N & \nodata  \\
4U~1916$-$053\tablenotemark{\dag} & 14 & N & \nodata \\
GX~17+2            & 12 & N & \nodata   \\
4U~1735$-$44       & 11 & N & \nodata  \\

\enddata
\tablenotetext{\dag}{There is a tentative detection of a burst
  oscillation frequency for this source, see Table \ref{tentsources}.}
\tablecomments{Most burst oscillations are seen when sources are in a
  particular accretion state with color-color diagram position
  variable $S_z>1.75$, (\citealt{Galloway08} and see the text for more detail).  This table lists all sources with more than 10 bursts in the RXTE Burst Catalogue (for bursts up to June 3 2007, \citealt{Galloway08}) for
  which burst oscillations have not been detected.  In most cases
  $S_z$ could not be computed since the source had not been observed
  over the full range of accretion states necessary to fully
  characterize the color-color diagram.  }
\end{deluxetable}

\subsubsection{Frequencies}
\label{frequencies}

The detection of burst oscillations in several persistent and
intermittent accretion-powered pulsars, sources for which the spin
frequency is known, has proven conclusively that burst oscillation
frequency is very close to the known spin frequency.  How close,
however, depends on the source. Table \ref{spinfreqs} summarizes the
situation for the various persistent and intermittent
accretion-powered pulsars with burst oscillations. In some cases the
frequencies agree to within $10^{-8}$ Hz, while for others they are
separated by a few Hz.  

\begin{deluxetable}{l p{4cm} p{8cm}}
\tabletypesize{\scriptsize}
\tablewidth{0pt}
\tablecaption{Comparison of burst oscillation and spin frequency.\label{spinfreqs}}
\tablehead{\colhead{Source}  & \colhead{References} &
  \colhead{Frequency comparison}   }
\tablecolumns{3}
\startdata
\sidehead{Persistent accretion-powered pulsars}
SAX J1808.4-3658 & \citet{Chakrabarty03} & Upwards drift of a few Hz
in the burst rise, starting below the spin frequency and
overshooting. In tail, frequency
exceeds spin by a few mHz. \\

IGR J17498-2921 & \citet{Chakraborty12} &  The frequency of burst oscillations in the tail (they are not detected significantly in the rise) appears to be relatively stable, and within $\pm 0.25$ Hz of the spin frequency.  However the precise offset, and limits on frequency drift, have yet to be quantified. \\

XTE J1814-338 & \citet{Strohmayer03}, \citet{Watts05}, \citet{Watts08b} & Frequency is
identical to the spin frequency within the errors ($\sim 10^{-8}$ Hz), with no drifts.  The
exception is the brightest burst, which shows a 0.1 Hz downwards drift 
in the burst rise.  \\

IGR J17511-3057 & \citet{Altamirano10a} & Upwards drifts of $\sim$ 0.1
Hz in the burst rise, starting below the spin frequency. There appears
to be overshoot in some bursts, but whether this is significant has
yet to be quantified. The frequency stabilises very close to the spin frequency in
the tail (certainly within 0.05 Hz), but the precise offset has also yet to be quantified.\\

IGR J17480-2446 & \citet{Cavecchi11} & Frequency is identical to the
spin frequency within the errors ($\sim 10^{-4}$ Hz), with no drifts.   \\

\sidehead{Intermittent accretion-powered pulsars}
Aql X-1 &  \citet{Zhang98}, \citet{Muno02a}, \citet{Casella08} &  Burst oscillations typically drift
upwards in frequency during the burst.  The maximum asymptotic
frequency reached by the burst oscillations is 0.5 Hz below the spin
frequency measured in the one brief episode where the source
showed accretion-powered pulsations. \\

HETE J1900.1-2455 & \citet{Watts09} & Burst oscillations drift upwards
by $\sim 1$ Hz during the burst, with the maximum frequency being
$\sim 1$ Hz below the spin frequency.  \\
\enddata
\end{deluxetable}

Most burst oscillation sources exhibit frequency drift, with the
frequency rising by a few Hz over the course of a burst.  Although we do not have sufficient counts to resolve individual cycles, the drift can be modelled and the resulting signals can be highly coherent, with values of Q as high as 4000 \citep{Strohmayer99}.    The most
comprehensive study of frequency evolution in burst oscillation trains
to date is that carried out by \citet{Muno02a}.  In most cases
the frequency drifts upwards during the burst, by 1-3 Hz, to reach an asymptotic
maximum (although the signal cannot always be tracked through
the peak of the burst since the amplitude sometimes drops below the
detectability threshold).  For the two intermittent pulsars, the
asympototic maximum is 0.5-1 Hz below the spin frequency (the drifts
for the regular pulsars are more unusual, see Table \ref{spinfreqs}).  The asymptotic maximum for a given source
appears to be very 
stable, with the fractional dispersion in asymptotic frequencies,
measured using bursts separated by several years, being $<
10^{-3}$. Although orbital motion might account for some of the
dispersion, it cannot account for all of the observed variation.    In
terms of coherence, the majority of the oscillation trains 
studied evolved smoothly in frequency, and hence appeared to be highly coherent.  However in about 30\% of cases, evolution was not smooth, suggesting jumps in phase or frequency,  or the simultaneous presence of two signals with very similar frequencies \citep{Muno02a}.  

The largest frequency drifts reported in the literature are of order a few Hz.  \citet{Wijnands01} reported an apparent drift of  5 Hz in a burst from X 1658-298, and  \citet{Galloway01} reported a drift of 3.6 Hz in a burst from 4U 1916-053.   In both cases, however, the burst oscillation signal drops below the detection threshold in the middle of the burst.  Since frequency cannot be tracked continuously throughout the burst, it is not clear whether we are really seeing a single signal drifting, or artificially connecting separate signals or perhaps even noise peaks (since with many bursts a few outliers due to noise would not be unexpected).  A rigorous analysis of the entire sample of burst oscillation observations, to determine limits on the size of the drifts that are compatible with the overall data set, has yet to be carried out.     

In the one case where oscillations were observed in a superburst, the frequency drift was much smaller (0.04 Hz
over 800 s), and entirely consistent with the expected Doppler shifts
due to orbital motion \citep{Strohmayer02a}.  The superburst oscillation frequency inferred from this
measurement was $\approx 0.4$ Hz higher than any of the 
aysmptotic frequencies measured during regular Type I X-ray bursts
from this source (see also \citealt{Giles02}).  If this source shows the
same behaviour as that seen in the intermittent pulsars
(Table \ref{spinfreqs}) it is likely that the superburst
oscillation frequency is closer to the true spin frequency of the star
than the frequency of the normal burst oscillations.

\subsubsection{Amplitudes}

The average amplitudes of burst oscillations (computed over the whole
burst) are highly variable, even for 
single sources, but are mostly in
the range 2-20\% rms 
\citep{Galloway08}.  Higher average amplitudes have been measured for
some bursts where the signal in the rise dominates (\citealt{Strohmayer97a},\citealt{Strohmayer98a},\citealt{Bhattacharyya06a}, \citealt{Galloway08}), although the error bars are larger since fewer photons can be
accumulated in the short burst rise.   For the oscillations detected during the superburst from 4U 1636-536, the amplitudes are $\sim 1$ \% rms, lower than that measured for oscillations detected in Type I bursts from this source \citep{Strohmayer02a}.

Amplitudes can also vary substantially during a burst. The
tendency for signals to disappear in the peaks of bursts has already
been mentioned.  This occurs most often, but not exclusively, during
periods of photospheric radius expansion \citep{Galloway08}, and is seen in both pulsars and
non-pulsars.   Significant variations in amplitude during bursts have
been found for a number of sources \citep{Muno02b} but there are
exceptions where the data are consistent with a constant amplitude
model \citep{Watts05}.  

For the sources that are accretion-powered pulsars, one can
compare the amplitudes of the burst oscillations to those of the
accretion-powered pulsations (\citealt{Chakrabarty03}, \citealt{Strohmayer03}, \citealt{Watts09},
  \citealt{Altamirano10a}, \citealt{Cavecchi11}, \citealt{Papitto11}, \citealt{Chakraborty12}).  The burst oscillation amplitudes for the
persistent pulsars are very different.  However for each
individual source the burst oscillation amplitude is within a few
percent
of (and mostly lower than) that the accretion-powered pulsations at
the time of the burst.   The difference between the amplitudes of the
two types of pulsations (for at least some bursts) is statistically
significant \citep{Watts05}.      

Harmonic content (a significant signal at the first
overtone of the burst oscillation frequency) has been found in most of the burst
oscillation trains from the persistent accretion-powered pulsars,
although not necessarily at the same level as that seen in the accretion-powered
pulsations (\citealt{Chakrabarty03}, \citealt{Strohmayer03}, \citealt{Watts05}, \citealt{Altamirano10a},
  \citealt{Cavecchi11}).  Burst oscillation only sources \citep{Muno02b}, and burst oscillations
from the intermittent pulsars (\citealt{Muno02b},\citealt{Watts09}), tend not to have significant harmonic
content, although see \citet{Bhattacharyya05b} for an
exception in the rising phases of some bursts.   

Another property that has been studied in some detail is the
dependence of burst oscillation amplitude on photon energy.  For burst
oscillation only sources, and burst oscillations from the intermittent
pulsars, amplitude rises with energy (\citealt{Muno03}, \citealt{Watts09}).
For persistent pulsars where this property has been analysed, however,
burst oscillation amplitude falls with energy
(\citealt{Chakrabarty03}, \citealt{Watts06}).  For sources that show
accretion-powered pulsations (persistent or intermittent) burst
oscillation amplitudes have the same energy dependence as the
accretion-powered pulsations, irrespective of whether this is a rise
or a fall (references above and \citealt{Casella08}).   

\subsubsection{Phase offsets}
\label{phases}

Searches for phase lags between pulse profiles constructed using
photons in different energy bands (energy dependent phase lags) have been carried out for
several burst oscillation sources. While in a few sources there are
marginal detections of hard lags (high energy pulse arriving later
than the low energy one), most burst oscillation profiles show no
statistically significant phase offsets (\citealt{Muno03}, \citealt{Watts06}, \citealt{Watts09}).  This
differs markedly from the behaviour of accretion-powered
pulsations, which have significant soft lags (\citealt{Cui98},
  \citealt{Gierlinski02}, \citealt{Galloway02}, \citealt{Kirsch04}, \citealt{Galloway05}, \citealt{Gierlinski05}, \citealt{Papitto10}).

For the persistent accretion-powered pulsars, one can also investigate
phase lags between the burst oscillation pulse profile and the
accretion-powered pulse profile.  For XTE J1814-338, which has no detectable
frequency drifts in its burst oscillations (for all but the
  final burst, which occurs at much lower accretion rates and has
  quite different properties), the two sets of pulsations are completely
phase-locked, with constant phase offset (\citealt{Strohmayer03}, \citealt{Watts08b}).  Indeed the burst oscillation profile is actually coincident (zero phase offset)
with the low energy accretion-powered pulse profile.  Spectral
modelling indicates that the latter component originates from the neutron star
surface (\citealt{Gierlinski02}, \citealt{Gierlinski05}).  Phase-locking of pulsations has also been reported
for IGR J17511-3057 \citep{Papitto10}, although frequency drifts in
this source complicate this type of analysis \citep{Altamirano10a}.

\section{Burst oscillation theories}

\subsection{Relevant physics}

Burst oscillations are associated with the surface
layers of the neutron star.  As such there are several key pieces of
physics that must be considered when developing models for the
phenomenon.  Before discussing specific models, it is worthwhile
reviewing these factors.  Some of these intrinsic properties remain
unchanged during an X-ray burst, whilst others may evolve.  

\subsubsection{Structure of the neutron star}

\begin{figure}
\begin{center}
\includegraphics[width=\columnwidth, clip]{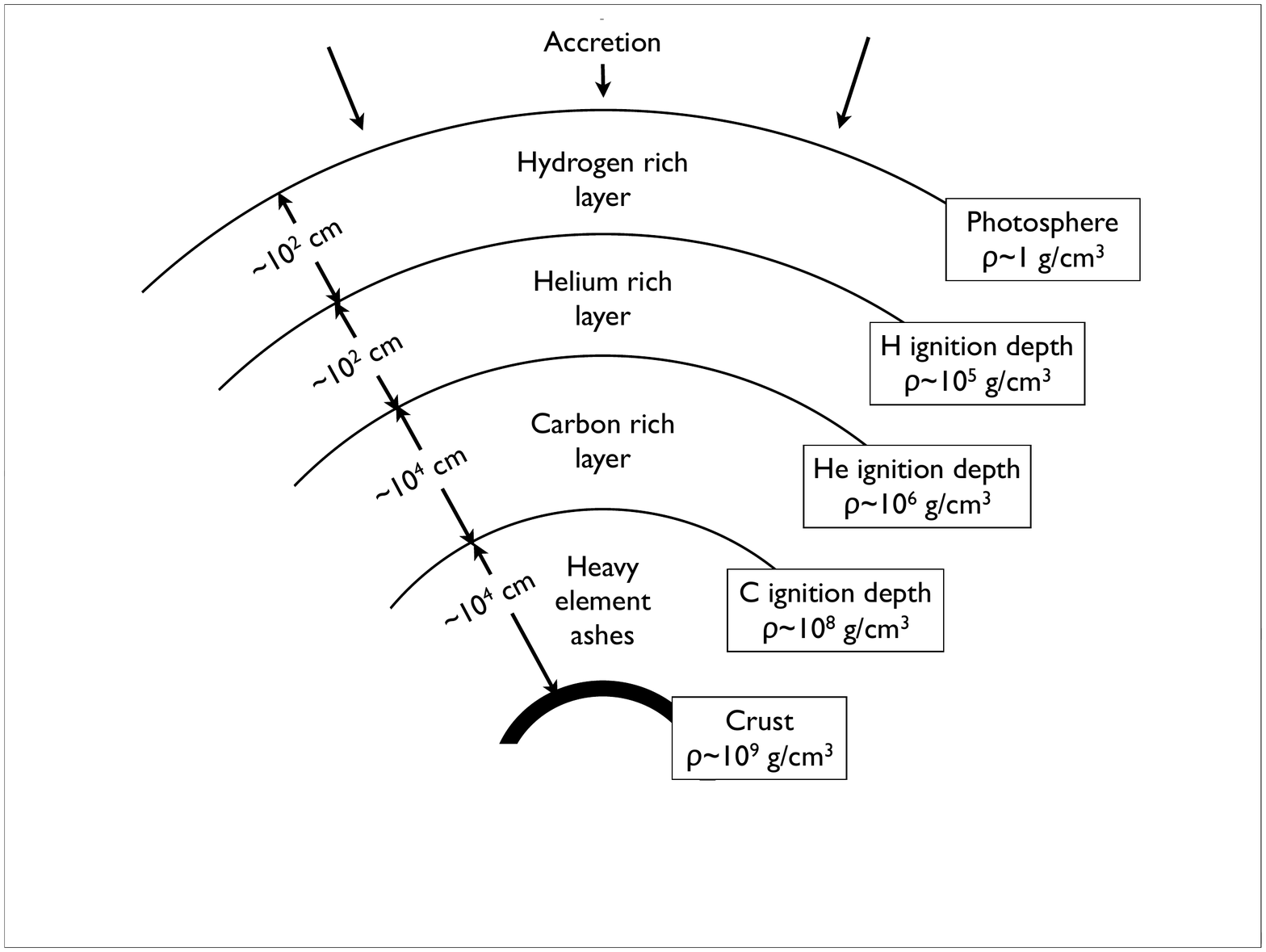}
\caption{The outer layers of an accreting neutron star, showing where thermonuclear
  ignition takes place.  Adapted from a figure in \citet{Lewin80}.  }
\label{oceanstructure}
\end{center}
\end{figure}

Figure \ref{oceanstructure} shows the stratified composition of the surface layers
of a neutron star, with ignition depths for different burst types
marked. When modelling how a burst develops, one needs
to take into account 
the coupling between the various burning and non-burning ocean layers, the
solid crust, and the photosphere. Coupling
may be dynamical, chemical, or thermal.  The layers will evolve and
expand due to heat generation during the burst, particularly during
episodes of strong photospheric radius expansion (\citealt{Paczynski83},
  \citealt{Paczynski86}, \citealt{intZand10}).

\subsubsection{Heat generation and transport}

Nuclear burning is unstable when the heat released by a 
themonuclear reaction causes an increase in reaction rate that cannot
be compensated for by cooling, resulting in a
thermonuclear runaway.  Several such reactions are involved in X-ray
bursts.  Hydrogen can burn unstably via the 
cold CNO cycle:  $^{12}\mathrm{C}(\mathrm{p}, \gamma)^{13}\mathrm{N}(\beta^+, \nu)^{13}\mathrm{C}(\mathrm{p},\gamma)^{14}\mathrm{N}(\mathrm{p}, \gamma)^{15}\mathrm{O}(\beta^+, \nu)^{15}\mathrm{N}(\mathrm{p}, \alpha)^{12}\mathrm{C}$, where the rate-controlling temperature dependent step is $^{14}\mathrm{N}(\mathrm{p}, \gamma)^{15}\mathrm{O}$.  Helium burns unstably via the temperature-dependent triple alpha reaction:  $3~^4\mathrm{He} \rightarrow ^{12}\mathrm{C} + \gamma$.  For reviews of the unstable reactions, and their application to X-ray bursts, see \citet{Schwarzschild65}, \citet{Hansen75}, \citet{Fujimoto81}, \citet{Lewin93}, and \citet{Bildsten98b}.  Superbursts are thought
to be caused by unstable burning of carbon (\citealt{Woosley76}, \citealt{Taam78},
  \citealt{Brown98}, \citealt{Cumming01}, \citealt{Strohmayer02b}). 

There are however uncertainties
inherent in modelling ignition and the progression of the
thermonuclear reactions that affect our understanding of the heat
generation process.  Ignition conditions are known to depend on accretion
rate, composition of accreted material, and heat flux from the deep
crust (all quantities that are impossible to measure precisely).
Burning, sedimentation, and mixing between bursts are also likely to play a role
(\citealt{Heger07}, \citealt{Peng07}, \citealt{Piro07}, \citealt{Keek09}).  Despite considerable theoretical
progress in this area, there still exists a substantial mismatch
between predicted and observed 
recurrence times for most bursters, indicating that our understanding of ignition remains
poor (\citealt{vanParadijs88}, \citealt{Cornelisse03}, \citealt{Cumming05b}, \citealt{Strohmayer06}, \citealt{Galloway08}), although there are a few exceptions (\citealt{Galloway06}, \citealt{Heger07b})   The precise
details of the reaction chains are also not exactly known, despite excellent
experimental work aimed at pinning this down \citep{Schatz11}. These uncertainties will
affect not only ignition conditions but also 
heat generation and compositional changes throughout the burst (see
for example \citealt{Fisker07}, \citealt{Cooper09}, \citealt{Davids11}, and \citealt{Schatz11}).  

Heat transport is also important in models of burst
oscillation development.  One-dimensional simulations show that radiative heat transport will be important
in all bursts, with convection also playing an important
role (\citealt{Woosley04}, \citealt{Weinberg06}).  Heat
transport will also be important in determining how the thermonuclear
flame spreads around the star.  Multi-dimensional calculations and
simulations have shown that convection, turbulence, conduction, and advection may
all play a role (\citealt{Fryxell82}, \citealt{Spitkovsky02}, \citealt{Malone11}).

\subsubsection{Rotation and flows}
\label{rotation}

If burst oscillation frequency is a good measure of spin frequency (as it appears to be, see Section \ref{frequencies}), then most burst oscillation sources rotate rapidly.  Is the rotation sufficient, however, to affect the evolution of the burst?
The importance of rotation in burst dynamics can be estimated by calculating the Rossby number
$R_o$.  This is the ratio of inertial to Coriolis
force terms in the Navier-Stokes equations, the hydrodynamical equations governing the
motions of the fluid layers where the burst takes place (see for
example \citealt{Pedlosky87}).  

\begin{equation}
R_o = \frac{U}{Lf}
\end{equation}
where $U$ is the velocity of the motion, $L$ a characteristic length,
and $f = 4\pi\nu_s\cos\theta$ the Coriolis parameter ($\nu_s$ being
the spin frequency of the star, and $\theta$ the co-latitude).
Rotation starts to have important dynamical effects once $R_o \lesssim
1$ and the Coriolis force becomes the dominant force. This means that rotational effects must be taken into
consideration once lengthscales exceed $U/f$.  As rotation rate
increases, the effects are felt on shorter lengthscales.  We can now estimate whether rotation is likely to affect burst 
oscillation mechanisms, given that oscillations have been found in stars
that rotate in the range 11-620 Hz (Table \ref{sources}). 

Let us first
consider the effect on flame spread.  One of the effects 
of the Coriolis force is to balance pressure 
gradients that develop as the hot fluid expands, hence slowing spreading \citep{Spitkovsky02}. The
appropriate speed for a flow driven by such pressure gradients is $U = \sqrt{gH}$ where $g$ is the
gravitational acceleration and $H$ the scale height of the hot fluid.  As rotation rate
increases, the tendency of the Coriolis force to confine the spreading
flame operates over ever shorter lengthscales.  Significant
confinement (slowing of flame spread) will occur once these are less
than the size of the star, $R$, 
\citep{Cavecchi11}.  This occurs for rotation rates

\begin{equation}
\nu_s > 25 ~\mathrm{Hz} \left(\frac{g}{10^{14}~ \mathrm{cm~s}^{-2}}
\right)^\frac{1}{2} \left(\frac{H}{10~\mathrm{m}}\right)^\frac{1}{2}
\left(\frac{10~\mathrm{km}}{R}\right)
\label{nusthresh}
\end{equation}
Clearly most of the burst oscillation sources are in the regime where
this will be relevant.

Rapid rotation will also affect oscillation modes (global wave patterns) that
might develop in the ocean and atmosphere, excited by the
thermonuclear burst.  As above, simple estimates can be used to
determine whether rotation will have an effect.  Shallow water gravity
waves (driven by buoyancy), for example, have a timescale $\tau = 1/N$, where $N =
\sqrt{g/H}$ is the Brunt-V\"ais\"al\"a frequency (see for example \citealt{Pringle07}).  The characteristic
speed of such waves is $U = H/\tau = \sqrt{gH}$.  This means that
modes with a wavelength $\lambda$ comparable to or larger than
$\sqrt{gH}/f$ will be sensitive to the 
effects of rotation.  Large-scale global modes (those
with the largest wavelengths,
$\lambda \sim R$) are
therefore likely to be affected by rotation for most of the burst
oscillation sources (compare to Equation \ref{nusthresh}).

Rotation can also affect ignition conditions.  Rotation lowers the effective gravity $g_\mathrm{eff}$ at the
equator as compared to the poles, with the difference being given by 
\begin{equation}
\frac{\delta g_\mathrm{eff}}{g_\mathrm{eff}} \sim
\left(\frac{\Omega_s}{\Omega_k}\right)^2 
\sim 5\%
\left(\frac{\nu_s}{500~\mathrm{Hz}}\right)^2
\left(\frac{R}{10~\mathrm{km}}\right)^3 \left(\frac{1.4~
    M_\odot}{M}\right),
\end{equation}
where $\Omega_k$ is the Keplerian angular velocity at the surface of the
neutron star.  The reduced gravity leads to a higher rate of fuel accumulation at the equator:   because of this ignition is likely to occur preferentially at equatorial latitudes for all but a small range of accretion rates (\citealt{Spitkovsky02}, \citealt{Cooper07d}).  

The expansion associated with the X-ray burst could also in principle
lead to differential rotation or shearing flows. Calculations show
that the change in thickness of the burning layers as they expand during a 
burst is $\Delta R \sim 20$ m \cite{Bildsten98b}.  If angular momentum
were conserved, this would lead to a spin down of order 

\begin{equation}
\Delta \nu_s \approx \frac{2\nu_s\Delta R}{R} = 2~\mathrm{Hz}
\left(\frac{\nu_s}{500~\mathrm{Hz}}\right) \left(\frac{\Delta
    R}{20~\mathrm{m}}\right) \left(\frac{10~\mathrm{km}}{R}\right)
\end{equation}
Over the course of the burst, the hot layer could in principle achieve
multiple wraps 
of the underlying star.  Zonal flows (latitudinal differential
rotation involving east-west flow along latitude lines, as opposed to meridional flows, which involve north-south flow
  along longitude lines) may also develop due to temperature
gradients as different 
latitudes ignite at different times \citep{Spitkovsky02}.  In practice any shear
flow (radial or latitudinal) would be attenuated by various
frictional processes or 
shearing instabilities, for instance.  However if shear flows do persist, they will
affect the hydrodynamics of the surface layers, modifying for example
the structure and stability of oscillation modes \citep{Pringle07}.

\subsubsection{Magnetic fields}
\label{mag}

Measuring magnetic field strength for the burst oscillation sources is very difficult.  For some of the pulsars, it has been possible to put limits on field strength by observing spin-down between accretion episodes and calculating the field strength necessary for this to occur via magnetic dipole radiation.  Using this method it has been determined that the accretion-powered pulsar SAX J1808.4-3658,
for example, has a field $\sim 10^8$ G (\citealt{Hartman08}, \citealt{Hartman09}). Most of the burst oscillation sources, however, are not pulsars.  If we assume that this is because the magnetic field is too weak to cause channeled accretion at the observed accretion rates, then this sets an upper limit on field strength $\sim 10^{9}$ G (for a
discussion of the techniques used in making this type of estimate see \citealt{Psaltis99}).  It is possible, however, that the magnetic field is simply aligned with the rotation axis \citep{Chen93} or that pulsations are obscured (\citealt{Titarchuk02}, \citealt{Gogus07}).  

The importance of magnetic fields on the
dynamics of the burning  
layers can be estimated by calculating the ratio of the Lorentz force
terms to the other force terms in the magnetohydrodynamic versions of
the Navier-Stokes equations.  The magnetic contribution is often split
into two terms - a magnetic pressure and a magnetic tension (see for
example \citealt{Choudhuri98}).  It should also be borne in mind that
the field is unlikely to remain static during the burst.  Flows set up by
the burst (convection, spreading of fuel or flame front, shearing
due to differential rotation) may lead to amplification and
modification of the field (\citealt{Cumming00}, \citealt{Boutloukos10}).  While magnetic pressure is unlikely
to be relevant compared to other pressure terms at the burning depth,
it will dominate in the very outermost layers of the atmosphere.  Magnetic
tension, which can act to counterbalance pressure gradients
associated with accumulating fuel and flame spread, may be dynamically important (\citealt{Brown98}, \citealt{Cavecchi11}).   Magnetic fields will also
introduce new types of oscillation and instability in the ocean layers
(see \citealt{Choudhuri98} for a general overview of magnetohydrodynamic oscillations, \citealt{Heng09}, and Section \ref{modes}).    Magnetic fields will also affect conditions in the ocean layers even
prior to ignition.  Channeling of accreting material onto the magnetic
poles of the neutron star may lead to a local over-density at the
magnetic polar caps, in addition to temperature and composition gradients \citep{Brown98}.  

\subsection{Current models for burst oscillations}

\begin{figure}
\begin{center}
\includegraphics[width=\columnwidth, clip]{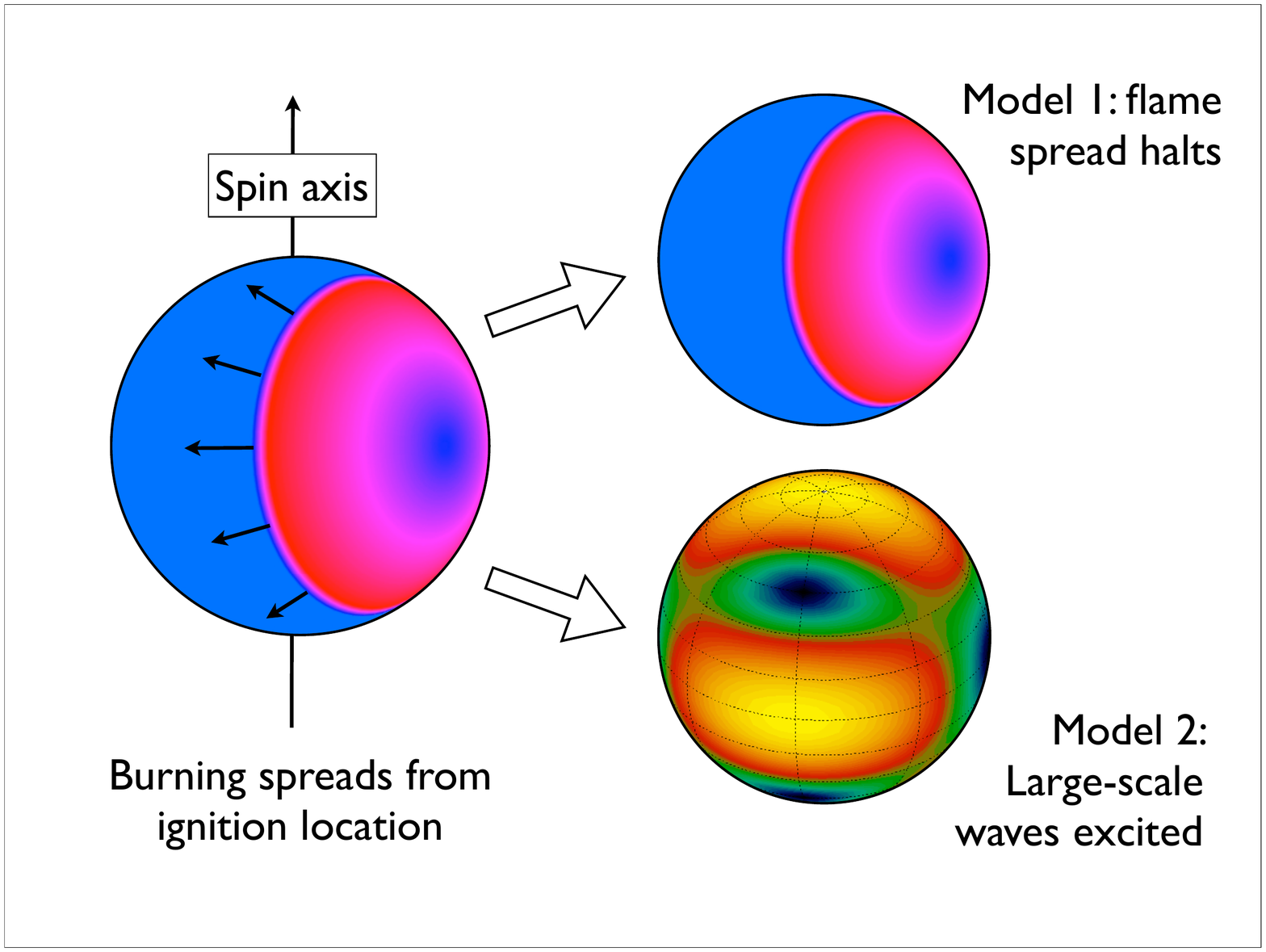}
\caption{Simplified schematic of current burst oscillation models.  Burst oscillations require the presence of a brightness asymmetry that can be modulated by the star's rotation so that the observer sees pulsations at close to the spin frequency.  Ignition is expected to start at a point, with a flame front then spreading across the star (left panel, red = hot, blue = cold).  This is known as a spreading hotspot.   What happens then is unclear.  One possibility is that the flame front stalls, so that only part of the ocean burns, leaving a hotspot that persists in the tail of the burst (top right).  The other possibility is that the flame front spreads around the star and excites large scale waves (global modes) in the ocean (bottom right).  The peaks and troughs of the waves will differ in temperature so that the wave pattern gives rise to a brightness asymmetry.}
\label{modelsum}
\end{center}
\end{figure}

Current models for burst oscillations fall into two categories, hotspot models and global mode models, illustrated in very general terms in Figure \ref{modelsum}.  As will be discussed in detail below, it seems likely that burning spreads initially from an ignition point, forming a spreading hotspot in the rising phase of the burst.  The burning front then either stalls, to leave a persistent hotspot, or spreads around the star, exciting large-scale waves in the ocean.  Both could lead to a brightness asymmetry in the ocean during the tail of the burst.  We will treat each class of model in turn, noting that they may not be mutually exclusive.    

\subsubsection{Hotspot models}
\label{hotspot}

Hotspot models are based on the general principle that the thermonuclear burning is not occurring uniformly across the surface of the star, but is instead confined to a smaller region.   This hotspot is then modulated by the star's rotation to give rise to the observed burst oscillations.    Two types of hotspot model have been considered: spreading hotspots, generated temporarily in the rise as the burst ignites and flame spreads from the ignition site to engulf the star; and persistent hotspots caused by restriction of the burning to a small part of the stellar surface.   

In the tails of most bursts, the blackbody radius obtained from fitting the overall burst spectrum is comparable to the radius of the star, implying that the flame has engulfed the entire surface (see for example \citealt{Galloway08}).   However it takes hours to days to accumulate fuel between bursts, and less than a second for a thermonuclear runaway to develop \citep{Shara82}.    For ignition conditions to be met simultaneously across the stellar surface, the thermal state would need to be consistent to better than one part in $\sim 10^4$ \citep{Cumming00}.  The presence of slight asymmetries in accretion (due for example to magnetic channeling, or the presence of equatorial boundary layers), makes this unlikely - particularly for the accretion-powered pulsars.   In the absence of a mechanism that could equalize the thermal state of the ocean to the required level, it is therefore assumed that ignition starts at a point and that a flame front then spreads across the stellar surface.  

The presence of such a spreading hotspot would provide a simple explanation for the presence of burst oscillations in the rising phase of bursts.  Results from a small sample of bursts appear to support this picture:  in a few bursts that have strong oscillations in the rising phase, the amplitude of oscillations falls as the blackbody radius of the overall burst spectrum increases (\citealt{Strohmayer97a}, \citealt{Strohmayer98a}, \citealt{Bhattacharyya06a}).  This is consistent with the idea of a flame front spreading to cover the star, since as the spot grows in size, the overall amplitude of pulsations will fall \citep{Muno02b}.    There are however several problems and theoretical questions that then arise.   

The first issue is that if point ignition and flame spread are occurring in all bursts, why do we not detect oscillations in the rising phase of all bursts?  As discussed in Section \ref{burstcond}, oscillations are actually more common in the tails rather than the rising phase of bursts.  One possibility is that this is a detection bias:  it can be difficult to make detections in the rising phase of bursts given their short duration.  Something that has yet to be done is a comprehensive study to check whether the current data, including upper limits, are consistent with the spreading hotspot model.

One factor that will also affect detectability is the latitude at which ignition occurs.  On a rapidly rotating star, ignition is expected near the equator for most accretion rates due to the lower effective gravity, which makes it easier to build up the column depth of material necessary for ignition \citep{Spitkovsky02}.  There are however small ranges of accretion rate close to the boundary between stable and unstable burning where ignition is expected to occur at higher latitudes \citep{Cooper07d}.  Whether a burst ignites at the equator or at higher latitudes can have a major effect on the detectability of burst oscillations \citep{Maurer08}.   For sources with a substantial amount of magnetic channeling, ignition may also occur preferentially at the magnetic poles due to local overdensities or extra heating \citep{Watts08b}.

How the flame then spreads is equally important to burst oscillation detectability, since the flame must be able to spread in such a way that an azimuthal asymmetry can persist.   The processes controlling flame spread in X-ray bursts have been an open question for years, with various heat transfer mechanisms including conduction, turbulence, and convection all thought to play a role (\citealt{Fryxell82},  \citealt{Nozakura84}, \citealt{Bildsten95b}, \citealt{Zingale01}).  \citet{Spitkovsky02} also pointed out the importance of hydrodynamical effects, in particular the interaction between the Coriolis force and expansion of the hot burning material.  The Coriolis force will act to slow down flame spread, thereby preserving a hotspot.  Indeed some degree of confinement may be essential to get the flame front established and propagating, since rapid expansion and lateral spread of the hot material might otherwise cause the flame to stall \citep{Zingale03}.   Note however that Coriolis force confinement will not be effective for IGR J17480-2446, which has a rotation rate of only 11 Hz (see Section \ref{rotation} and \citealt{Cavecchi11}).  The flame spread mechanism remains an unsolved problem with immense importance for burst oscillation models.   

If the flame spreads across the whole star during the rising phase, the spreading hotspot will dissipate.  The fact that different regions of the star ignite at different times will leave some residual temperature asymmetry in the tail.  However the magnitude of the predicted difference is too low to explain the amplitudes of the burst oscillations observed in burst tails \citep{Cumming00}.  Some additional mechanism, or modification to the model, is still required.   One possibility is the excitation of large-scale waves (Section \ref{modes}).  The other is that a hotspot survives because the burning front does not spread across the whole star.

One way in which this might occur is if fuel is confined, with the prime candidate for a confinement mechanism being the star's magnetic field.  Matter channeled onto the magnetic polar caps of the star will be prevented from crossing field lines (and hence flowing across the star) until the overpressure is sufficiently large to deform the field lines.  Estimates by \citet{Brown98} show that for matter to remain confined in the ignition depth for normal Type I bursts requires fields of at least  $10^{10}$ G, much higher than the fields strengths estimated for the burst oscillation sources (Section \ref{mag}).  Magnetohydrodynamical instabilities may also act to make magnetic confinement of fuel ineffective \citep{Litwin01}.  Magnetic fuel confinement on burst oscillation sources is therefore expected to be relatively unlikely.

The alternative is that the flame front itself is confined:  burning, once started, spreads some distance and then stalls.   This requires heat transport to be impeded in some way.   Several mechanisms have been explored in the literature.   \citet{Payne06} suggested that magnetic field evolution on an accreting neutron star might result in the development of strong magnetic belts around the equator that would then impede north-south heat transport.   On rapidly rotating stars, a reduction in Coriolis-mediated confinement as the flame front tries to cross the equator might result in stalling (\citealt{Spitkovsky02}, \citealt{Zingale03}).  Both of these mechanisms would confine burning to one hemisphere of the star: it is unclear whether this would lead to an asymmetry of sufficient magnitude to explain burst oscillation amplitudes.    Recently \citet{Cavecchi11} showed that the dynamical interaction between the expanding burning fuel and the radial magnetic field could in principle induce horizontal field components that might be large enough to restrict further spread.  This would lead to more localized confinement.  For this mechanism to work the initial field can in principle be somewhat lower than for fuel confinement, $\sim 10^9$ G (consistent with estimates for two sources, XTE J1814-338 and IGR J17480-2446).   If ignition starts at the magnetic pole, this would result in burst oscillations that are phase-locked to the accretion-powered pulsations and have only minimal frequency drift, consistent with observations from these two sources (Sections \ref{frequencies} and \ref{phases}).   This mechanism may not work, however, for sources that appear to have lower magnetic fields, unless something in the burst process (such as a convective dynamo) acts to increase the field temporarily \citep{Boutloukos10}.  

A major question addressed even in the earliest papers on burst oscillations was how to explain the upwards frequency drift seen in most bursts (Section \ref{frequencies}).  \citet{Strohmayer97a} pointed out that horizontal motion of the spreading flame front could lead to drifts in frequency of a few Hz.  However in this case rotation would have to set a preferred direction for spread, to explain why the frequency always tends to rise.  They therefore suggested an alternative possibility: that of angular momentum conservation during vertical expansion of the heated burning layers.  Rapid expansion would cause a hotspot to rotate more slowly, giving an observed frequency below the spin rate, with frequency rising as the layers cooled and contracted again (see Section \ref{rotation}).  

The expansion model depends on two key assumptions:  that the burning layers can decouple from the underlying star; and that shearing within the burning layer itself is small enough to preserve the spot rather than causing it to be smeared out.   The validity of these assumptions, and the magnitude of the expected frequency shift in first Newtonian gravity and then General Relativity, were examined by \citet{Cumming00} and \citet{Cumming02}.  These authors examined a range of hydrodynamical coupling mechanisms, such as viscous or shearing instabilities, and found that it was in principle possible for the burning layer to remain decoupled for the timescale of the burst.   Smearing out of the hotspot was more problematic, and keeping this coherent required convection or short wavelength baroclinic instabilities to operate inside the burning layer.  Smearing was found to be particularly pronounced during episodes of photospheric radius expansion.  The maximum frequency drifts predicted by these studies were $\sim 1-2$ Hz.  Although larger drifts have been reported, as explained in Section \ref{frequencies} the largest drifts in the literature are not clear cut, and the data set as a whole may in fact be consistent with a smaller maximum drift.     Magnetic wind-up and shearing effects may also be important in determining the coupling and angular momentum transfer between the expanding layers and the underlying star (\citealt{Cumming00}, \citealt{Lovelace07}).  A role for the magnetic field may explain the different drift properties seen in burst oscillations from the accretion-powered millisecond pulsars, where the magnetic field is known to be dynamically important.

\subsubsection{Surface modes}
\label{modes}

Ignition of the burst and the subsequent spread of flame around the star may excite large-scale waves (global modes) in the neutron star ocean.   Height differences associated with such oscillations would translate into hotter and cooler patches (the wave pattern), thus giving rise to differences in X-ray brightness.   Non-axisymmetric modes could in principle be excited easily by the initial flame spread, and persist throughout the burst tail (depending on damping mechanisms).  

The problem of finding global modes of oscillation in oceans is one with a long and venerable history, due to its applications to the Earth's oceans.    As such, papers on this topic use a rich and sometimes confusing mix of terminology when referring to different mode types.    The basic hydrodynamical equations of continuity (mass conservation), momentum and energy, applied to the oscillations of a shallow stably stratified ocean on a rotating sphere, reduce to what is known as Laplace's tidal equation (\citealt{Longuet68}, \citealt{Townsend05}, \citealt{Lou00}).   The restoring forces in this simple system are buoyancy (due to temperature or composition gradients) and the Coriolis force. 

Non-axisymmetric mode solutions to Laplace's tidal equation have dependence $\exp(im\varphi)$ on the azimuthal angle $\varphi$, where $m$ is an integer.   Given a mode with an azimuthal number $m$ and frequency $\nu_r$ in the rotating frame of the star, the frequency $\nu_o$ seen by an inertial observer would be 

\begin{equation}
\nu_o = m\nu_s + \nu_r
\label{obsfreq}
\end{equation}
where $\nu_s$ is the spin frequency of the star, and the sign of $\nu_r$ is positive or negative depending on whether the mode is prograde (eastbound) or retrograde (westbound) respectively.  The angular structure of the modes, which depend on spherical harmonics, are characterized by the usual integers $l$ and $m$ (see for example \citealt{Mathews70}).  The pattern has $m$ nodes in azimuth and $|l-m|$ nodes in latitude.   For non-zero $\nu_r$,  the brightness pattern moves around the star, permitting an offset from the spin-frequency.  Frequency drift can be naturally explained if $\nu_r$ depends on the thermal state of the ocean, since this will change as the burst evolves and cools through the tail.  

Laplace's tidal equation admits three families of mode solution:  

\begin{itemize}
\item{Poincar\'{e} modes:   propagate in the retrograde (westbound) direction and reduce to pure gravity waves (restored by buoyancy alone) in the non-rotating limit.    For this reason they are often referred to as g-modes in the neutron star ocean oscillation literature (see for example \citealt{Bildsten96} and  \citealt{Heyl04}). }
\item{Kelvin modes:  propagate in the prograde (eastbound) direction and also reduce to pure gravity waves in the non-rotating limit.  Unlike the Poincar\'e modes, they are in geostrophic balance.  This means that the horizontal component of the Coriolis force balances the horizontal pressure gradient, see \citet{Pedlosky87}. As a result they involve purely azimuthal motions, and are confined to an equatorial waveguide.}
\item{Rossby modes:  propagate in the retrograde (westbound) direction, and are also known as r-modes  (\citealt{Papaloizou78a}, \citealt{Saio82}).  In the limit of zero buoyancy, Rossby modes are driven purely by latitudinal variations in the Coriolis force.  In the non-rotating limit they reduce to trivial zero-frequency solutions.  If buoyancy is present then it will also act to restore these modes, and they are then often referred to as buoyant r-modes (see for example \citealt{Heyl04} and \citealt{Piro05b}).}
\end{itemize}

\citet{Heyl04} was the first to give serious consideration to whether such ocean modes might explain the properties of burst oscillations, particularly in the tails of bursts.  He laid out a number of criteria that the modes would have to meet, including the fact that observed frequency should be a few Hz below the spin frequency of the star, and that low $l, m$ modes would be needed to ensure that a highly sinusoidal and high amplitude burst oscillation would result (the finer the pattern, the lower the contrast across the stellar surface).   On the basis of these criteria he determined that the $l=2, m=1$ buoyant r-mode was the most likely candidate.  This retrograde mode has $|\nu_r|$ of only a few Hz, giving an observed frequency slightly below the spin frequency.  As the burning layers cool (changing the buoyancy), $|\nu_r|$ gets smaller, so that observed frequency would tend towards the spin frequency.   Poincar\'e or g-modes were ruled out primarily on the grounds of their frequencies, since $|\nu_r|$ is expected to be much larger than a few Hz (\citealt{McDermott87}, \citealt{Strohmayer96a}, \citealt{Bildsten95a}, \citealt{Bildsten96}, \citealt{Bildsten98c}, \citealt{Heyl04}, \citealt{Maniopoulou04}). Kelvin modes were ruled out on the grounds that they are prograde, so that frequency would fall as the ocean cooled, in contrast to the majority of observed frequency drifts.  Both Kelvin and Poincar\'e modes are also strongly confined to the equatorial regions on rapidly-rotating stars, which would reduce overall burst oscillation amplitudes (since there is no strong contrasting mode pattern across most of the surface).   The buoyant r-mode, by contrast, occupies a wider band around the equator, giving greater contrast.

In addition to providing a natural explanation for the sinusoidal shape, overall frequency, and amplitude of the burst oscillations, mode models predict that the amplitude of the burst oscillation should increase with photon energy (\citealt{Heyl05}, \citealt{Lee05}, \citealt{Piro06}), in line with observations of the non-pulsars and the intermittent pulsars (although not those of the persistent pulsars).   \citet{Heyl04} also argued that excitation of such a mode should be relatively easy, since the timescale of the mode matches the $\sim 1$ s rise time of the burst associated with the flame spread.   \citet{Narayan07} and \citet{Cooper08} explored the issue of excitation in more detail, concluding that the mode would need to be unstable in order to grow to the required amplitudes in the tails of bursts.  They considered various classical pumping mechanisms \citep{Saio93} and conclude that the main driver is likely to be the $\epsilon$ mechanism, which drives modes via nuclear energy generation.  The role of the $\epsilon$ mechanism in neutron star envelopes was studied previously by \citet{McDermott87} and \citet{Strohmayer96a} for thermally unstable envelopes.  \citet{Narayan07} and \citet{Cooper08}, by contrast, consider thermally stable envelopes, appropriate for the tails of bursts.   For He-rich explosions,  instability is more likely when the burst is radiation pressure dominated.  Instability was much less likely to develop in H-rich bursts.   Suppression of the instability occurs during episodes of convection, which may contribute to the observed disappearance of burst oscillations during the peaks of bright photospheric radius expansion bursts.  

The major problem with the buoyant r-mode model is that the frequency drift predicted as the burning layer cools is $\sim 10$ Hz, an order of magnitude larger than the observed values.    At $10^9$ K, a typical temperature in the peak of a burst, the buoyant r-mode frequency should be $|\nu_r| \gtrsim 10$ Hz, drifting to $|\nu_r| \sim 1$ Hz as the layer cools to $10^8$ K.     Various solutions to this problem have been explored.  \citet{Piro05a} pointed out the existence of another global mode of oscillation that is confined to the interface between the cool ocean below the burning layer and the underlying elastic crust.  The observed frequency of this crust-interface mode would be $\sim 4$ Hz below the spin frequency of the star.  If, as the buoyant r-mode drifted upwards in frequency, it could couple efficiently to this crustal-interface mode and transfer its energy via resonant conversion, this would naturally truncate the drift a few Hz short of the spin frequency \citep{Piro05b}.  The frequency of the crustal interface mode is highly stable, so could account for the reported invariance in asymptotic frequencies.  It could also explain why the burst oscillation frequency seen in a superburst is close to that seen in normal bursts, despite the differences in the burning layer, since the final frequency in both cases would be set by the crust-interface mode.   Unfortunately detailed studies of the coupling between the buoyant r-mode and the crustal-interface wave indicate that resonant conversion is not effective \citep{Berkhout08}.  The fact that the separation between spin frequency and burst oscillation frequency is much less than 4 Hz for the persistent and accretion-powered pulsars (Table \ref{spinfreqs}) is also problematic for the crust-interface mode model.

As a result, several alternative models have been explored.  \citet{Heyl04} pointed out that if the r-mode were instead confined to the photospheric layers, the drift would be smaller.   This idea has yet to be explored in detail, but as pointed out by \citet{Berkhout08} magnetic forces will become important in these low density layers, and could have a substantial effect on mode frequencies.    \citet{Cumming05a} explored what might happen to the buoyant r-modes if strong zonal flows (latitudinal differential rotation) were to develop in the ocean layers in the immediate aftermath of flame spread.  While shear-modified modes could in principle match both frequencies and drifts, his study showed that high $m$ modes were likely to grow much faster than the low $m$ modes necessary to explain burst oscillation properties.  Whether zonal flows can develop in the first place is also still an open question \citep{Spitkovsky02}.      Most recently \cite{Heng09} have considered the effects that magnetic fields might have on oceanic mode structure, by solving the shallow water magnetohydrodynamical equations for an incompressible ocean with a purely radial magnetic field.  In addition to magnetically modified versions of the normal families of modes described above,  they find new solutions dominated by magnetic and rotational effects that they call magnetostrophic modes.   Unlike the previously known modes, which are confined to the equatorial regions on a rapidly rotating star, these new modes are stronger near the poles.   Both prograde and retrograde modes are possible:  however the model predicts a downwards rather than an upwards frequency drift for the retrograde modes as the ocean cools.   More detailed calculations, however, are required to determine whether or not magnetically modified modes such as these might play a role in explaining burst oscillations.  

Mode models remain an extremely promising mechanism to explain burst oscillations from the non-pulsars and the intermittent pulsars, although issues such as the frequency drift problem remain to be resolved.   For the persistent accretion-powered pulsars, they have more difficulty:  existing mode models cannot explain the extremely close agreement between spin and burst oscillation frequency (Table \ref{spinfreqs}), which would require $|\nu_r| \ll 1$ Hz for an $m=1$ mode; the unusual frequency drifts (or lack thereof); and the fact that burst oscillation amplitude decreases with photon energy.  Magnetically modified modes may eventually account for these differences:  alternatively sources with a dynamically important magnetic field may have a different burst oscillation mechanism (such as a magnetically confined hotspot, see Section \ref{hotspot}).  

\section{Uses of burst oscillations}

Burst oscillations are an intrinsic part of the thermonuclear burst process.  They can also, however, be used as tools to explore other areas of neutron star physics.  In this Section I will expand briefly on some of these applications.  

\subsection{Pulse profile modelling as a diagnostic of mass and radius}

The nature of the strong force, and the state of matter at extremes of temperature and density, is an extremely active field of research.   Of particular interest is the transition from nucleons (neutrons and protons) to de-confined quarks and gluons or other more exotic states. At low densities this can be studied by experiments like the Large Hadron Collider (LHC) or in various Heavy Ion experiments such as the Relativistic Heavy Ion Collider (RHIC) and the Facility for Antiproton and Ion Research (FAIR).  At higher densities, however, neutron stars are the only environment in the Universe where the transition can be explored \citep{Paerels10}. 

Dense matter models, combined with the relativistic stellar structure equations, make predictions for mass and radii of neutron stars.   By measuring these quantities, we can thus work back to the underlying nuclear physics \citep{Lattimer07}.  Burst oscillations are expected to be powerful tools in this regard because the emission comes from the surface of the neutron star, deep within the gravitational potential well.  As photons propagate towards us, they are affected by relativistic effects that depend on mass and radius, such as gravitational redshift and light-bending.  The pulse profile associated with the surface emission pattern is modified by these processes.  By modelling the shape and energy-dependence of the observed pulse profile using relativistic ray-tracing algorithms, researchers seek to deconvolve these effects and hence extract information about mass and radius.   Typically one has to fit for the surface emission pattern and the geometry (such as observer inclination), as well as the parameters of interest.

The fact that stellar compactness will affect pulse profiles from a hotspot on the surface of the neutron star is long-established, with a series of papers exploring the effects of (for example) different metrics, rotation, and stellar oblateness (\citealt{Pechenick83}, \citealt{Chen89}, \citealt{Strohmayer92}, \citealt{Braje00}, \citealt{Cadeau05}, \citealt{Cadeau07}, \citealt{Morsink07}).  The techniques developed in these papers are generic to all types of hotspot and have been applied extensively to accretion-powered pulsations as well as burst oscillations.   Following the discovery of burst oscillations, several authors started to model the lightcurves and pulse profiles that might be expected from both static and spreading hotspots or mode patterns on the surface of bursting neutron stars (\citealt{Strohmayer97a}, \citealt{Miller98}, \citealt{Weinberg01}, \citealt{Muno02b}, \citealt{Muno03}, \citealt{Viironen04}, \citealt{Lee05}).    Some attempts were also made at the inverse problem:  that of fitting measured burst oscillation pulse profiles in order to obtain confidence contours on stellar compactness (\citealt{Nath02}, \citealt{Bhattacharyya05a}).  Many more attempts have been made to obtain constraints by fitting accretion-powered pulsations, despite the complicating effects of the accretion funnel and shock on the emission, see \citet{Poutanen08} for a review.  Burst oscillations offer in principle a somewhat cleaner system dominated by thermal emission.   Although the constraints obtained using existing data were relatively weak, the potential of the technique to put tight bounds on mass and radius, given sufficient photons, was clear \citep{Strohmayer04}.  As such pulse profile modelling of burst oscillations is now a major science driver for proposed future large area X-ray timing missions such as the {\it Large Observatory for X-ray Timing} (LOFT, \citealt{Feroci11}) and the {\it Advanced X-ray Timing Array} (AXTAR, \citealt{Ray10}).   A concerted effort is now underway to address and resolve some of the outstanding issues that may introduce degeneracies into fits for mass and radius, such as the effect of changes in the surface emission pattern over the course of a burst.  

\subsection{Neutron star spin}

The evidence outlined in Section \ref{frequencies} supports the identification of burst oscillation frequency as a good measure of neutron star spin, at least to within a few Hz.   The main consequence of this has been to double  the number of rapidly-rotating (above 10 Hz) accreting neutron stars with a known spin rate.  Indeed the fastest spinning accreting neutron star currently known (4U 1608-522, with a spin of 620 Hz) is a burst oscillation source and not an accretion-powered pulsar.    

Identifying the spin distribution of the various classes of neutron
star has been a longstanding research goal for many years, as they are an important element in our understanding of stellar and binary evolution.  The discovery of rapidly rotating accreting neutron stars, in particular, was regarded as critical to confirming the recycling scenario for the formation of the millisecond radio pulsars (\citealt{Alpar82}, \citealt{Radhakrishnan82}, \citealt{Bhattacharya91}).  With a larger number of stars in this class more detailed studies of population evolution are possible, revealing interesting discrepancies between models and observation (\citealt{Tauris06}, \citealt{Lorimer08}, \citealt{Hessels08}, \citealt{Kiziltan09}, \citealt{Tauris11}).    These point to problems in our understanding of, for example, the mass transfer process, magnetic field decay, and accretion torques.    Other formation routes for millisecond radio pulsars, such as accretion-induced collapse of white dwarves, may also need to be invoked to explain the observed populations.  

The maximum spin rate that a neutron star can reach is also of great interest.  A neutron star with a sub-millisecond spin period would place a tight and very clean constraint on the dense matter equation of state (\citealt{Lattimer07}, \citealt{Schaffner08}, \citealt{Haensel09}).    This is because the break-up, or mass-shedding, frequency (the spin rate above which a neutron star would fly apart as rotational forces overwhelm gravitational attraction) depends strongly on the composition of the star.   This in turn depends on the poorly constrained behavior of the strong force at supranuclear densities.   Although no current burst oscillation source (and indeed not even the most rapidly rotating radio pulsar, \citealt{Hessels06}) rotates fast enough to rule out equations of state, the search for extremely rapid rotators does motivate burst oscillation searches.  Whether burst oscillations could ever develop, or be detectable, on such rapidly rotating stars remains however an open question (see Section \ref{rotation}).  

The current spin distribution of accreting neutron stars, as given by both the accretion-powered pulsars and the burst oscillation sources, hints at a maximum that is well below the spin rate suggested by simple estimates of accretion-induced spin-up over their lifetime (\citealt{Chakrabarty03}, \citealt{Chakrabarty08}).  If the evolutionary estimates
are sound, there is a requirement for a braking mechanism to halt the
spin-up.   Magnetic braking (due to interaction between the stellar
field and the accretion disk) is one possibility (\citealt{Ghosh78}, \citealt{White97}, \citealt{Andersson05}).  Another, which
has generated a lot of excitement, is the
emission of gravitational waves (\citealt{Papaloizou78b}, \citealt{Wagoner84}, \citealt{Bildsten98a}).   This has triggered major theoretical effort exploring the nature of possible gravitational wave generation mechanisms, such as crustal mountains (\citealt{Bildsten98a}, \citealt{Ushomirsky00}, \citealt{Melatos05}, \citealt{Haskell06}), or internal r-mode oscillations (\citealt{Bildsten98a}, \citealt{Andersson99}, \citealt{Levin99}, \citealt{Andersson00}, \citealt{Andersson02}, \citealt{Heyl02}, \citealt{Wagoner02}, \citealt{Nayyar06}, \citealt{Bondarescu07}, \citealt{Ho11}).    Gravitational wave braking could also make accreting neutron stars promising continuous wave sources for future gravitational wave detectors, although
this is one application where knowing the spin to a very high
degree of precision (certainly better than a few Hz) would be important \citep{Watts08a}.  Uncertainty in the spin rate has a very negative impact on gravitational wave detection threshold, since increasing the parameter space increases the numbers of trials (reducing detection sensitivity, see Section \ref{detection}) and increases the computational power needed to perform any search.    Improving burst oscillation models so that we can make a more accurate diagnosis of spin rate would be of great benefit.

\section{Conclusions}

Burst oscillations are an intriguing part of the phenomenology of thermonuclear bursts on neutron stars, with wider uses in terms of measuring stellar spin and potentially constraining the dense matter equation of state.  One of the legacies of the Rossi X-ray Timing Explorer has been a rich database of burst oscillation observations, giving us a clear understanding of their general properties and the conditions under which they develop.  A satisfactory theoretical explanation of why and under what conditions burst oscillations develop, however, is still lacking.   Burst evolution models that take full account of both the structure of the neutron star ocean and envelope, and the dynamical effects of magnetic fields, are needed to explore the mechanics of flame spread and the excitation of global oceanic modes.  Such theoretical development is essential if we are take full advantage of the opportunities offered by the next generation of X-ray timing telescopes.  

\section*{Acknowledgments}

ALW acknowledges support from a Netherlands Organisation for Scientific Research (NWO) Vidi Fellowship.   She would like to thank Michiel van der Klis and Deepto Chakrabarty for their encouragement and feedback whilst writing this review.  She would also like to thank Jean in 't Zand, Duncan Galloway, Diego Altamirano, Tullio Bagnoli, Yuri Levin and Yuri Cavecchi for their comments on the draft manuscript.

\bibliographystyle{Astronomy}

\bibliography{araaburst}   % if natbib is available

\end{document}